\UseRawInputEncoding
\documentclass{IEEEtran}
\pdfoutput=1
\usepackage{algorithmic}
\usepackage{graphicx}
\usepackage{cite}
\usepackage{amsmath,amssymb,amsfonts}
\usepackage{algorithmic}

\usepackage{textcomp}
\usepackage{xcolor}

\usepackage{algorithm}
\usepackage{lineno,hyperref}
\usepackage{multirow}
\usepackage{epstopdf}

\usepackage{float}
\usepackage{caption}

\usepackage{hyperref}

\usepackage{color,array,amsthm}

\newtheorem{property}{Property}

\newtheorem{remark}{Remark}
\newtheorem{assumption}{Assumption}
\def\BibTeX{{\rm B\kern-.05em{\sc i\kern-.025em b}\kern-.08em
		T\kern-.1667em\lower.7ex\hbox{E}\kern-.125emX}}
\begin{document}

\title{Spacecraft Attitude Pointing Control under Pointing Forbidden Constraints with Guaranteed Accuracy \\
}

\author{Jiakun Lei,
	Tao Meng,
	Weijia Wang,
	Shujian Sun,
	Heng Li 
	Zhonghe Jin
	\thanks{Jiakun Lei, Ph.D. School of Aeronautics and Astronautics, Zhejiang University, Hangzhou, China, 310027, 12124010@zju.edu.cn}
	
	\thanks{Tao Meng, Prof., School of Aeronautics and Astronautics, Zhejiang University, Hangzhou, China, 310027; Zhejiang Key Laboratory of micro-nano-satellite, mengtao@zju.edu.cn}

	\thanks{Weijia Wang, Ph.D., School of Aeronautics and Astronautics, Zhejiang University, Hangzhou, China, 310027, 12024055@zju.edu.cn}
	
		\thanks{Shujian Sun, Prof., School of Aeronautics and Astronautics, Zhejiang University, Hangzhou, China, 310027, Zhejiang Key Laboratory of micro-nano-satellite, sunshujian@zju.edu.cn}
		
			\thanks{Heng Li, M.S., Zhejiang Key Laboratory of micro-nano-satellite, Hangzhou, China, 310027, hengli0911@outlook.com}
	
	\thanks{Zhonghe Jin, Prof., School of Aeronautics and Astronautics, Zhejiang University, Hangzhou, China, 310027, Zhejiang Key Laboratory of micro-nano-satellite, jinzh@zju.edu.cn}
}

\maketitle

\begin{abstract}
This paper focuses on the attitude pointing control problem under pointing-forbidden constraints and performance constraints. The spacecraft is expected to align its sensor's boresight to a desired direction, while the terminal control accuracy and the attitude adjustment rapidity should also be guaranteed simultaneously.
To resolve this problem, a switching controller structure is proposed in this paper based on the reduced-attitude representation, fusing the artificial potential field (APF) methodology and the Prescribed Performance Control (PPC) scheme together. 
Firstly, a novel artificial potential field is presented, and a particular function is designed for the mollification of the switching process, aiming at providing a smooth transition for the system status.
Subsequently, we propose a special performance function, which can freeze the PPC part when necessary. In this way, the intrinsic contradictory between the fast attitude maneuver and forbidden direction avoidance is tackled
 Further, an asynchronous switching strategy is designed, guarantees the system's stability. Based on these proposed issues, a switching backstepping controller is developed, and a tracking differentiator(TD) is employed to generate a smooth approximation of differential signals. Numerical simulation results are illustrated to show the effectiveness of the proposed scheme.
\end{abstract}


\section{Introduction}
For the real on-orbit attitude pointing control mission, celestial bodies always exist that the sensor's boresight should circumvent, which motivates the requisition for the obstacle-avoidance attitude controller. On the other hand, since contemporary space missions are presented with many additional requirements, the performance requirements for the attitude controller are also placed in a vital position recently. Motivated by this topic, this paper investigated the attitude pointing control problem under pointing constraints, while the performance issues are also taken into consideration. The spacecraft is supposed to align its boresight vector to the desired direction, while the control accuracy and the rapidity should be also guaranteed.

 \textbf{1. Pointing-Constrained Attitude Control} 
 
 Attitude control under pointing constraints is one of the most widely discussed topics, and its existing solutions can be mainly classified into two types: the one based on planning and the one without planning.
For the planning-based solution, it often relies on an advance planning of a constrained problem, as stated in \cite{biggs2016geometric,he2022pointing,xu2018rotational,wu2017time,wu2019energy,xu2017rapid}. Although these planning-based control methods are remarkable, it often requires complex computing, which increasing the computational burden of the on-board computer. Among the ones without planning, the most considered is the control based on the artificial potential field technique(APF). There's many APF proposed in the existing literature, as presented in \cite{hu2019anti,chi2018reduced,feng2019reorientation}. In \cite{bai2021torque}, an APF considering limited controller output is presented, and dynamic obstacles are well circumvented. In \cite{dongare2021attitude}, based on a reduced-attitude representation, a bounded smooth repulsion field with an explicit reaction region is designed, and the practical-asymptotically convergence of the system is guaranteed. Also, parameter regulation can guarantee that these obstacles will not be reached. In a word, these APF-based controllers provide an economical way to handle the pointing constraint, which motivates our topic.

 \textbf{2. Performance-Constrained Attitude Control} 
 
 For the performance-constrained attitude control problem, the prescribed performance control (PPC) scheme is often adopted, which is a unique inspiration work proposed by Bechlioulis and Rovathakis in \cite{bechlioulis_adaptive_2009}. The main idea of PPC is to transform the state-constrained control problem into an equivalently unconstrained one. Since the original state variable is constrained by the performance requirements, through a kind of homeomorphic mapping, the original constrained error variable can be translated to an unconstrained one, on which the controller can be developed directly. Another way to realize the PPC is to employ the Barrier Lyapunov function (BLF) technique, as stated in \cite{wei2021overview}. In PPC, the state trajectory is forced to stay in a sector-like region enclosed by the prescribed performance function. Since the performance function is a monotonically decreasing one related to the performance requirements, the given requirements can be satisfied by ensuring the state trajectory stays in the constraint region during the whole control procedure. As listed in \cite{shao2018fault,hu2018adaptive,wei2018learning,amrr2021prescribed,luo2018low,zhang2019prescribed}, the PPC scheme has been introduced to the attitude control problem of spacecraft to show its effectiveness.

 \textbf{3. Attitude Control with Performance Constraints and Pointing Constraints} 
 
For an efficient way to handle the attitude pointing problem with performance constraints and pointing constraints, it is still a lack. Intuitively, this problem may could be solved through planning and closed-loop control. However, this may result in a high computational cost. Another natural idea is to fuse the prescribed performance control (PPC) scheme and the artificial potential function (APF) technique together. Although this seems a rational solution, we found some key issues that make the direct combination of PPC and APF hard to perform: 

$1.$ There is an intrinsic contradiction between the PPC scheme and the APF scheme when the system meets obstacles. For the PPC-governed controller, the state trajectory will converge rapidly to the steady-state in order to satisfy the performance requirements. For the APF-governed controller, the system will definitely slow down its convergence rate when meets obstacles, ensuring that the system is able to avoid the obstacle region.
Therefore, when the system is working under PPC scheme and APF scheme simultaneously, since PPC will providing a fast convergence of the system even when there exist obstacles, the system may not be able to circumvent the obstacle under such a high convergence rate, causing the violation of the pointing constraints. In this way, the directly mixing of PPC scheme and APF scheme cannot guarantee the system's stability.

$2. $ The general representation of the attitude pointing constraint expressed in the unit quaternion or Modified Rodrigues Parameter(MRP) is a quadratic constraint, which makes it hard to incorporate with the PPC scheme directly. Owing to this reason, the pointing constraint is hard to directly incorporated into the typical PPC error transformation.

In order to conquer these two main issues in this problem, we come up with the following idea: since the performance requirements can be tackled through PPC scheme, and the attitude-pointing constraint can be solved through APF scheme, the system can be designed to be a switching one. The system will working under PPC scheme if the system is far away from the obstacles, while the system will switches to APF when system meets obstacles. Notably, there exists a time interval during the system transition progress such that the system will work under PPC scheme and APF scheme together. To ensure the system's stability, we design a mechanism to freeze the PPC scheme under this condition.
Since the PPC system is governed by the state variable and the performance function simultaneously, we noticed that the PPC scheme can be freeze by changing the performance function correspondingly. Thus, the PPC's translated variable remain unchanged under such a condition, and the system Lyapunov function will be governed only by the APF part. Therefore, the intrinsic contradiction can be solved in this way. 

Further, we found that the pointing constraint will be simple under the reduced-attitude representation. To elaborate this problem, we made following analysis: in order to guarantee that the translated variable is able to exponentially-converged, a virtual control law can be derived accordingly. Meanwhile, the attraction field of the APF will exert attractive effect to the system, ensuring the convergence of the pointing error. When compare these two effect, we found that its derived virtual angular velocity has the same direction, which indicates the virtual control law derived from PPC and the virtual control law derived from APF's attraction field can be integrated together. Such a characteristic plays a significant role in this paper as it allows the direct combination between PPC and APF under reduced-attitude representation.

Following these stated ideas, a switching controller that fusing the PPC scheme and the APF scheme together is presented to handle the attitude pointing control problem with performance requirements and pointing constraints.
The main contribution of this paper can be mainly concluded as follows:

\textbf{1.}In order to tackle the intrinsic contradiction between APF and PPC when the system meets obstacles, motivated by \cite{yong2020flexible}, a special prescribed performance function (PPF) governed by a switching strategy is designed for the system's stability. The performance function will be exponential-decayed when the system is far away from the obstacles, ensuring the rapid convergence of the system. On the other hand, the performance function will change accordingly to hold an unchanged translated variable when the system meets obstacles, freezing the PPC system and guarantee that the system is able to circumvent the exclusion zone. In this way, the aforementioned intrinsic contradiction between PPC and APF scheme is well tackled.

\textbf{2.}  A newly-designed differentiable piece-wise function is proposed to mollify the switching progress, providing a smooth system switching process. Further, the proposed function is designed with explicitly parameter design, and the transition process of switching can be designed easily.

\textbf{3.} 
Owing to the application of the reduced-attitude representation, the incorporation of the PPC scheme and the APF scheme is realized for the attitude control problem.
Based on the reduced-attitude representation, the attitude pointing control problem with pointing constraints and performance constraints are resolved without using planning-based methods. A complete control scheme with a switching structure is proposed for this topic.

\section{Preliminaries}
\subsection{Notations}
In this paper, we define the following notations for analysis.
$\|\cdot\|$ stands for the Euclidean norm of a vector or the induced norm of a matrix, while the operator $\boldsymbol{r}^{\times}$ represents the $3 \times 3$ skew-symmetric matrix for vector cross manipulation, i.e. $\boldsymbol{r}^{\times}\boldsymbol{s} = \boldsymbol{r}\times\boldsymbol{s}$. $\text{ang}\left(\boldsymbol{a},\boldsymbol{b}\right)$ denotes the angle between vector $\boldsymbol{a}$ and $\boldsymbol{b}$, where $\boldsymbol{a}$, $\boldsymbol{b}$ are all unit vectors. The body-fixed frame is denoted as $\mathfrak{R}_{b}$, while the earth central-fixed frame is denoted as $\mathfrak{R}_{i}$ correspondingly.

\section{Problem Formulation}
\subsection{Reduced Attitude Representation}
In this paper, the reduced-attitude representation is employed.
Define a boresight vector of the sensor fixed on the spacecraft as $\boldsymbol{B}_{b}$, and define the  pointing-direction expressed in the inertial frame as $\boldsymbol{r}_{i}$. Denote the corresponding expression of $\boldsymbol{r}_{i}$ expressed in the body-fixed frame as $\boldsymbol{r}_{b}$, we can define the following variable for the evaluation of the pointing error, expressed as follows \cite{dongare2021attitude}:
\begin{equation}
	   x_{e} = 1 - \boldsymbol{B}_{b}^{\text{T}}\boldsymbol{r}_{b} = 1 - \cos\left(\text{ang}\left(\boldsymbol{B}_{b},\boldsymbol{r}_{b}\right)\right)
\end{equation}
Obviously, if the sensor's boresight vector is aligned with the desired direction, we have $x_{e} = 0$;
 Notably, since  $\text{ang}\left(\boldsymbol{B}_{b},\boldsymbol{r}_{b}\right)\in\left[0,\pi\right]$ is hold, $x_{e}\in\left[0,2\right]$ will be always satisfied. For $\text{ang}\left(\boldsymbol{B}_{b},\boldsymbol{r}_{b}\right) = 0$, $x_{e} = 0$ will be hold; for $\text{ang}\left(\boldsymbol{B}_{b},\boldsymbol{r}_{b}\right) = \pi$, we have $x_{e} = 2$.
Take the time-derivative of $x_{e}$, the kinematics and the dynamics of the reduced attitude system can be arranged into the following form:
\begin{equation}
	\label{system}
	   \begin{aligned}
	   	\dot{x}_{e} &= -\boldsymbol{B}^{\text{T}}_{b}\boldsymbol{r}^{\times}_{b}\boldsymbol{\omega}_{s}\\
	   	\boldsymbol{J}\dot{\boldsymbol{\omega}}_{s} &= -\boldsymbol{\omega}^{\times}_{s}\boldsymbol{J}\boldsymbol{\omega}_{s} + \boldsymbol{u} + \boldsymbol{d} 
	   \end{aligned}
\end{equation}
where $\boldsymbol{J}\in\mathbb{R}^{3}$ represents the inertial matrix of the spacecraft expressed in the body-fixed frame $\mathfrak{R}_{b}$. $\boldsymbol{u}$ represents the control input of the system, and the lumped external disturbance is denoted by $\boldsymbol{d}$.
\begin{remark}
	For the attitude-reduced representation, notably, we find that there exists two equilibrium point such that $\dot{x}_{e} = 0$. It can be discovered that $x_{e} = 0$ is a stable equilibrium point of the system, while $x_{e} = 2$ is the unstable equilibrium point of the system.
\end{remark}
For the synthesize of the proposed control scheme, we made the following assumption in this paper.
\begin{assumption}
	\label{ASS1}
	The lumped external disturbance is unknown but bounded by a known constant, denoted as $D_{m}$. This indicates that for arbitrary time instant, we have $\|\boldsymbol{d}\| \le D_{m}$.
\end{assumption}
\begin{assumption}
	{\label{assump_J}}
	The inertial matrix $\boldsymbol{J}$ is a known symmetric positive-definite matrix. Thus, we have:
	\begin{equation}
		\lambda (\boldsymbol{J})_{\text{min}}\boldsymbol{x}^{\text{T}}\boldsymbol{x} \le \boldsymbol{x}^{\text{T}}\boldsymbol{J}\boldsymbol{x} \le
		\lambda (\boldsymbol{J})_{\text{max}}\boldsymbol{x}^{\text{T}}\boldsymbol{x}
	\end{equation}
	where $\lambda\left(\cdot\right)$ represents the corresponding eigen value of the matrix $\boldsymbol{J}$. $\lambda_{J}\left(\cdot\right)_{\text{max}}$ denotes the maximum eigen value of $\boldsymbol{J}$, while the minimum eigen value of $\boldsymbol{J}$ is denoted as $\lambda_{J}\left(\cdot\right)_{\text{min}}$.
\end{assumption} 

\subsection{Pointing-Forbidden Constraint}
\begin{figure}[hbt!]
	\centering 
	\includegraphics[scale = 0.4]{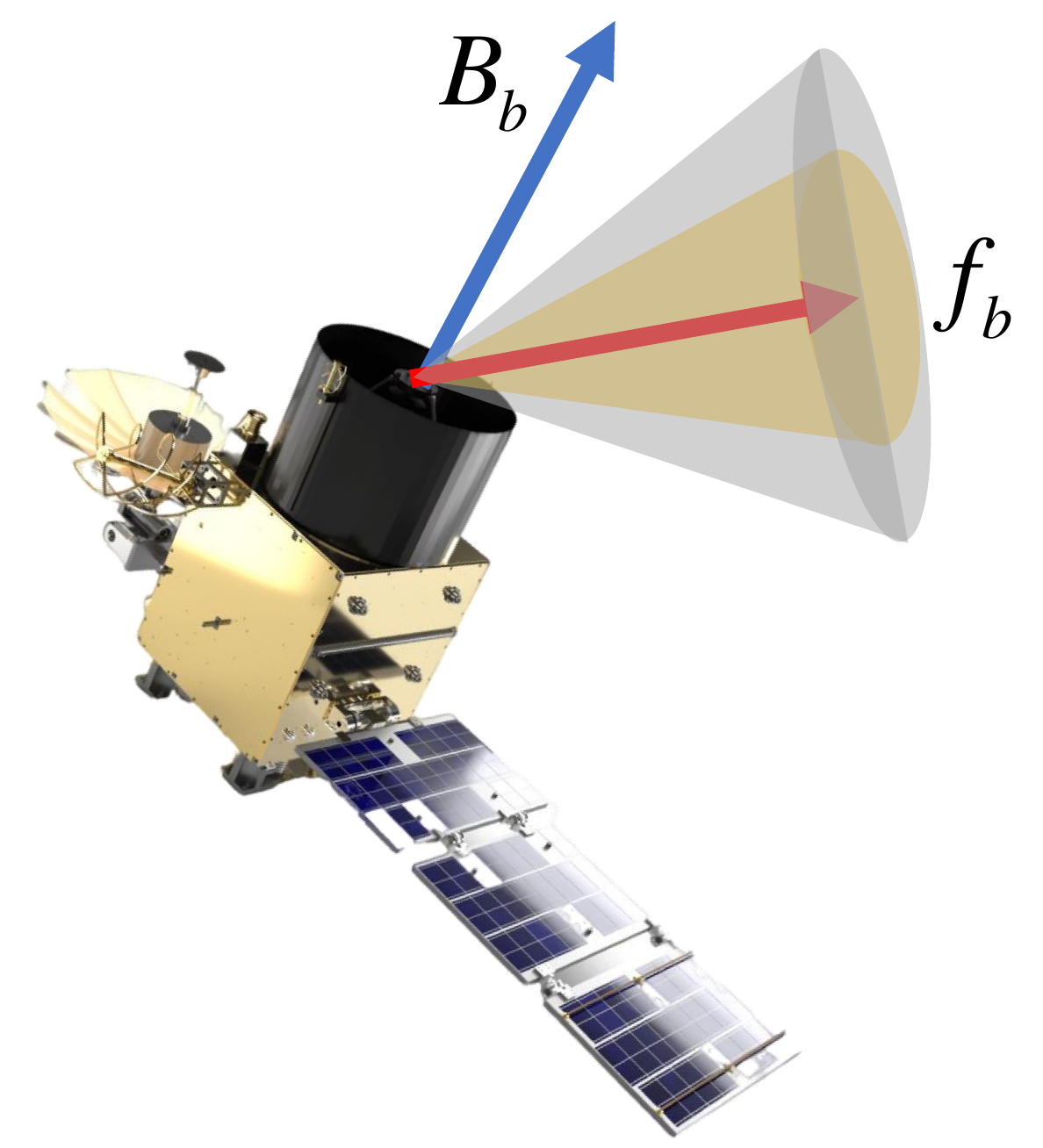}
	\caption{Illustration of spacecraft's pointing constraint}       
	\label{Sat}   
\end{figure}

For the safety consideration, there exists some directions that the sensor's boresight vector should circumvent. Define the restricted direction vector expressed in the inertial frame as $\boldsymbol{f}_{i}$, its expression in the body-fixed frame can be written as $\boldsymbol{f}_{b} = \boldsymbol{A}_{bi}\cdot\boldsymbol{f}_{i}$. Accordingly, the pointing-forbidden constraint can be established as follows:
\begin{equation}
	     \cos\left(\text{ang}\left(\boldsymbol{B}_{b},\boldsymbol{f}_{b}\right)\right) = \boldsymbol{B}^{\text{T}}_{b}\boldsymbol{f}_{b} \le \cos\Theta_{f}
\end{equation}
where $\Theta_{f}$ denotes the minimum forbidden angle for the pointing constraint. For instance, $\Theta_{f} = 20^{\circ}$ is considered in this paper. A sketch map of the pointing constraint is illustrated in Figure [\ref{Sat}].

In Figure [\ref{Sat}], the yellow cone represents the pointing-forbidden region such that $\boldsymbol{B}_{b}$ should never invade it.

\section{Control Objective}
This paper aims to develop a control scheme to ensure that the pointing constraint and the control accuracy requirements can satisfy simultaneously. The system should be able to ensure that the sensor's boresight vector can circumvent all the obstacles in the space, and the system should remain at a rapid convergence rate to the steady-state, with a guaranteed pointing control accuracy achieved. Besides, all the closed-loop signals should be bounded by a small value, and the system should be able to converge practical-asymptotically. 

\section{Problem Solution}

This paper employs the PPC scheme and the APF technique to handle the performance requirement and the pointing constraint, fusing these two schemes together. The system is designed to be a switching system, and will switching automatically when meets obstacles. A special switching strategy for prescribed performance function (PPF) and an asynchronous switching strategy is designed for the system's stability. Besides, the tracking differentiator (TD) technique is employed in the control scheme to generate the virtual control law's differential signal, ensuring the tracking to the virtual control law.

The following solution of the system is organized as follows: the main procedure of the PPC part such as the error transformation, Barrier Lyapunov function design and the prescribed performance function design is presented in subsection \ref{errtrans}, \ref{BLF} and \ref{PPFdesign}, respectively. Further, the detailed elaboration of APF designing is elaborated in subsection \ref{APFdesign}. Elaboration of the system switching strategy will be stated in section\ref{SW}, while the statements about the tracking differentiator  will be mentioned in subsection \ref{TD}.

\subsection{Error Transformation Procedure}
\label{errtrans}
As typical PPC scheme stated\cite{wei2021overview}, the state trajectory should stay in a constraint region enclosed by the prescribed performance function(PPF). Denote arbitrary PPF as $\rho$, the state constraint can be established as
$-\rho\left(t\right) < e\left(t\right) < \rho\left(t\right)$.
Further, define a translated error variable as $\varepsilon\left(t\right) = e\left(t\right)/\rho\left(t\right)$, the given state constraint can be rearranged into an equivalently form:
\begin{equation}
	\label{Cons}
	|\varepsilon\left(t\right)| < 1
\end{equation}
Therefore, once the constraint in equation (\ref{Cons}) is satisfied, the performance requirements will be able to achieve.

Define the PPC translated variable of $x_{e}$ as $\varepsilon_{q}$, and further denote the corresponding performance function of $\varepsilon_{q}$ as $\rho_{q}$. Taking the time-derivative of $\varepsilon_{q}$, one can be obtained that:
\begin{equation}
	\label{deq}
	\dot{\varepsilon}_{q} = \frac{1}{\rho_{q}}\left[\dot{x}_{e}-\frac{\dot{\rho}_{q}}{\rho_{q}}x_{e}\right]
\end{equation}

\subsection{Barrier Lyapunov Function Design}
\label{BLF}
In order to guarantee that the constraint (\ref{Cons}) is able to be satisfied all along, we employ the Barrier Lyapunov Function (BLF) methodology to handle the performance constraint. Based on our previous work \cite{WANG2022}, the BLF is designed as follows in this paper.
\begin{equation}
	V_B = gF\ln\left[\cosh\left(\varepsilon/F\right)\right]
\end{equation}
where $g > 0$ and $F > 0$ are design parameters. Unlike those mostly applied "Logarithimic" type BLF, the designed BLF will be meaningful for arbitrary $\varepsilon$, avoiding the potential singularity of traditional BLF functions, as stated in \cite{WANG2022}.

 \subsection{Switching Prescribed Performance Function(S-PPF)}
 \label{PPFdesign}
 In this paper, the prescribed performance function(PPF) is a special one generated by a switching system, expressed as follows:
 \begin{equation}
 	\label{dRho}
 	\dot{\rho} = -k_{\rho}\left(\rho-\rho_{\infty}\right)\left(1-\Omega_{s}\right)+\left(\dot{e}/e\right)\rho\Omega_{s}
 \end{equation}
where $\rho_{\infty}$ stands for the terminal value of PPF, $k_{\rho} > 0$ controls the decaying rate of the PPF. $\Omega_{s}$ is the switching parameter, of which the detail will be elaborated later in section \ref{SW}.  The initial condition of this system is defined as $\rho\left(0\right) = \rho_{0}$.

Considering about the given PPF, the working status for $\rho$ can be mainly divided into two cases.

\textbf{Mode 1.} For $\Omega_{s} = 0$, the PPF system can be written as $\dot{\rho} = -k_{\rho}\left(\rho-\rho_{\infty}\right)$, hence the PPF is an exponential converged one under this condition. It can be observed that $\rho\to\rho_{\infty}$ will be hold for $t\to+\infty$ under this case, which indicates that $\rho_{\infty}$ is an asymptote of the PPF.

 \textbf{Mode 2.} For $\Omega_{s} = 1$, the system will be governed by $\dot{\rho} = \left(\dot{e}/e\right)\rho$. Notably, since $\dot{\varepsilon} = d\left(e/\rho\right)/dt$, it can be obtained that $\dot{\varepsilon} = \frac{\dot{e}\rho - e\dot{\rho}}{\rho^{2}} = 0$ will be hold for current condition, which indicates that the translated variable $\varepsilon$ will remain unchanged in this case.

 For $\Omega_{s}\in\left(0,1\right)$, it can be inferred that $\rho$ will be governed by these two parts simultaneously, and the specific condition of how $\rho$ changed is depend on the dominant one. Since $\Omega_{s}$ is governed by a smooth differentiable function, it can be derived that $\rho$ will be always differentiable.
 f
 \subsection{Artificial Potential Field Design}
 \label{APFdesign}
 In this paper, motivated by the work in \cite{dongare2021attitude}, we design the following APF function. The attraction field $U_{a}$ is defined as follows:
 \begin{equation}
	U_{a} = k_{a}x_{e}
\end{equation}
where $k_{a} > 0$ stands for the gain of the attraction field. It can be observed that $U_{a}$ is a proportional one related to $x_{e}$.

For the repulsion field, it is constructed based on a designed smooth piece-wise function. Define $\beta = \boldsymbol{B}^{\text{T}}_{b}\boldsymbol{f}_{b} = \cos\left(\text{ang}\left(\boldsymbol{B}_{b},\boldsymbol{f}_{b}\right)\right)$, the repulsion field can be expressed as follows:
\begin{equation}
	\label{UB}
	U_{r} = \begin{cases}
		0 &  \beta\in\left[-1,L_{0}\right)\\
	\frac{k_{r}}{2}\left[\tanh\frac{r\left(L_{1}-L_{0}\right)/k_{r}\cdot\left(\beta-L_{m}\right)}{\sqrt{(\beta-L_{0})(L_{1}-\beta)}}+1\right] & \beta\in\left[L_{0},L_{1}\right)\\
		k_{r} &  \beta\in\left[L_{1},1 \right]\\
	\end{cases}
\end{equation}  
where $k_{r}>0$ stands for the gain of the repulsion field. $L_{0} = \cos\Theta_{0}$ and $L_{1} = \cos\Theta_{1}$ are angular parameters related to the pointing constraint. Notably, the repulsion field will start working from $\beta \ge L_{0}$, and $\beta\in\left(L_{0},L_{1}\right)$ can be regarded as a buffer zone of the system. Accordingly, $\Theta_{0}>\Theta_{1}$, $\Theta_{1}\ge\Theta_{f}$ should be satisfied. $r$ is related to the maximum slope of $U_{r}$. Let $L_{m} = \frac{1}{2}\left[L_{0} + L_{1}\right]$, we have $\partial{U}_{r}/\partial \beta = r$ for $\beta = L_{m}$. This provides an explicitly design of the repulsion field's gradient.

\begin{assumption}
	\label{ASS2}
	There's no repulsion field at the desired pointing direciton, 
	and the minimum angle between the desired pointing direction $\boldsymbol{r}_{b}$ and the pointing-forbidden vector $\boldsymbol{f}_{b}$ satisfies $\min\left(\text{ang}\left(\boldsymbol{r}_{b},\boldsymbol{f}_{b}\right)\right) \ge \Theta_{df}$, where $\Theta_{df}$ is a constant. In this paper, $\Theta_{df} = 50^{\circ}$ is considered.

\end{assumption}
	This assumption indicates that the forbidden direction will not be too close to the desired direction. Such an assumption is rational for a real constrained attitude maneuver scenario.

\begin{figure}[hbt!]
	\centering 
	\includegraphics[scale = 0.8]{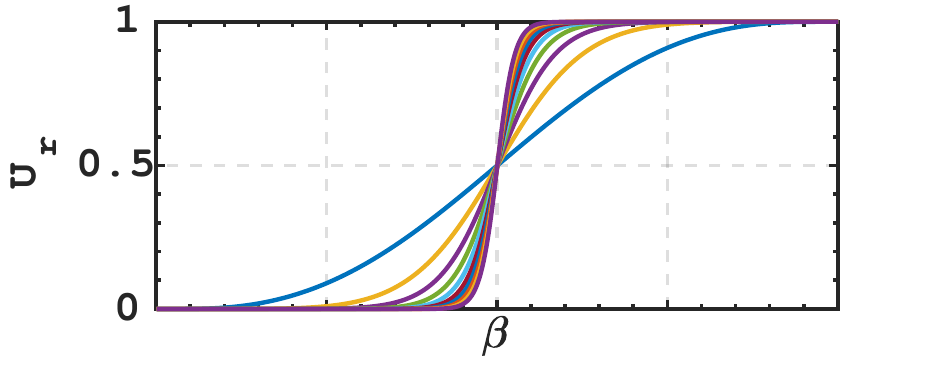}
	\caption{Repulsion field under different $r$}      
	\label{URUR}     
\end{figure}

A sketch map of the designed repulsion field under different $r$ is illustrated in Figure [\ref{URUR}], where $r$ from $1$ to $10$ is considered sequentially.
	It can be discovered that the bigger $r$ is, the steeper the function is. However, it will also result in a decreasing of the valid effecting region of $U_{r}$, as $U_{r}$ will be close to $0$ for most of the time. Hence, the selecting of $r$ should be set appropriately to hold continuously repulsion force to the state trajectory.
	
Based on the proposed $U_{a}$ and $U_{r}$, the total APF is designed to be $U = U_{a} + U_{r}$. For multiple pointing constraints, the total APF is established as $U = U_{a} + \sum U_{ri}$, where $U_{ri}$ represents the $i$ th repulsion field exerted by the exclusion zone.

For convenience, considering that there exists only one pointing forbidden region, take the time-derivative of $U_{a}$, one can be obtained:
\begin{equation}
	\label{dUa}
	\dot{U}_{a} = -k_{a}\boldsymbol{B}^{\text{T}}_{b}\boldsymbol{r}^{\times}_{b}\boldsymbol{\omega}_{s}
\end{equation}
Take the time-derivative of $U_{r}$, one can be obtained that:
\begin{equation}
	\label{dUr}
	\dot{U}_{r} = 
		k_{r}\nabla_{\beta}\boldsymbol{B}^{\text{T}}_{b}\boldsymbol{f}^{\times}_{b}\boldsymbol{\omega}_{s}
\end{equation}
where $\nabla_{\beta}$ stands for the partial derivative of $U_{r}/k_{r}$ w.r.t. $\beta$. Therefore, we have $k_{r}\nabla_{\beta} \le r$

\begin{remark}
	Note that the given potential function may not a convex one for some condition, however, owing to the application of the reduced-attitude representation, a simple strategy to avoid the singularity can be considered in this scheme, as stated in \cite{guo2011spacecraft}. On the other hand, the strict proof of the APF's convexity is not the main issue of our paper.
\end{remark}
\subsection{Tracking Differentiator}
\label{TD}
In this paper, we employ the tracking differentiator (TD) stated in \cite{yang2019active} to generate differential signals for the virtual control law. Define the filter output as $\boldsymbol{x}_{1}$, $\boldsymbol{x}_{2}$, the employed TD can be expressed as follows:
\begin{equation}
	\begin{aligned}
		\dot{\boldsymbol{x}}_{1} &= \boldsymbol{x}_{2}\\
		\dot{\boldsymbol{x}}_{2} &= -R^{2}\left[\tanh\left(\boldsymbol{x}_{1}-\boldsymbol{u}\right)\right]\\
		&\quad-R^{2}\left(a_{2}\left(\tanh\left(\boldsymbol{x}_{2}/R\right)\right)\right)
 	\end{aligned}
\end{equation}
where $a_{1}$, $a_{2}$, $R$ are design parameters.
It has been proved in \cite{yang2019active} that the tracking differentiator will asymptotically converge to the origin point, thus all the closed loop error will be bounded such that $\|\boldsymbol{x}_{1}-\boldsymbol{u}\|\le \epsilon_{1}$, $\|\boldsymbol{x}_{2}-\dot{\boldsymbol{u}}\|\le \epsilon_{2}$, where $\epsilon_{1}$ and $\epsilon_{2}$ are small constants.
\begin{remark}
	As stated in section \ref{TD}, the presented tracking diffentiator dosen't require the differentiable of the input signal, which is suitable for our switching controller.
\end{remark}
\section{ System Switching Strategy}
\label{SW}
In this paper, we design an asynchronous switching strategy for the system. The system will be governed by two switching variable denoted as $\Omega_{s}$ and $\Omega_{v}$, stated as follows:

\begin{equation}
	\label{Omegas}
	\Omega_{s} = \begin{cases}
		0 &  \beta\in\left[-1,V_{0}\right)\\
		\frac{1}{2}\left[\tanh\frac{m\left(V_{1}-V_{0}\right)\cdot\left(\beta-V_{m}\right)}{\sqrt{(\beta-V_{0})(V_{1}-\beta)}} + 1\right] & \beta\in\left[V_{0},V_{1}\right)\\
		1 &  \beta\in\left[V_{1},1 \right]\\
	\end{cases}
\end{equation}  

\begin{equation}
	\Omega_{v} = \begin{cases}
		0 &  \beta\in\left[-1,P_{0}\right)\\
		\frac{1}{2}\left[\tanh\frac{n\left(P_{1}-P_{0}\right)\cdot\left(\beta-P_{m}\right)}{\sqrt{(\beta-P_{0})(P_{1}-\beta)}} + 1\right] & \beta\in\left[P_{0},P_{1}\right)\\
		1 &  \beta\in\left[P_{1},1 \right]\\
	\end{cases}
\end{equation} 
where $m$, $n$ control the increasing rate of $\Omega_{s}$ and $\Omega_{v}$ respectively.
In this section, we design the transition process of $\Omega_{s}$ to be an aggressive one, while the switching of $\Omega_{v}$ is designed to be a mild one.

Here we give some main principles of the switching strategy's parameter selecting.
\textbf{1.}	For $\Omega_{s}$, $V_{0} = L_{0}-2\delta$, $V_{1} = L_{0}$ and $V_{m} = L_{0} - \delta$ are considered, with $\delta$ is a small positive value.
\textbf{2.}	Different from $\Omega_{s}$, we suggest that $P_{0} = V_{1}$, $P_{1}\le L_{1}$, $P_{m} = \frac{1}{2}\left[P_{0}+P_{1}\right]$.

\begin{figure}[hbt!]
	\centering 
	\includegraphics[scale = 0.9]{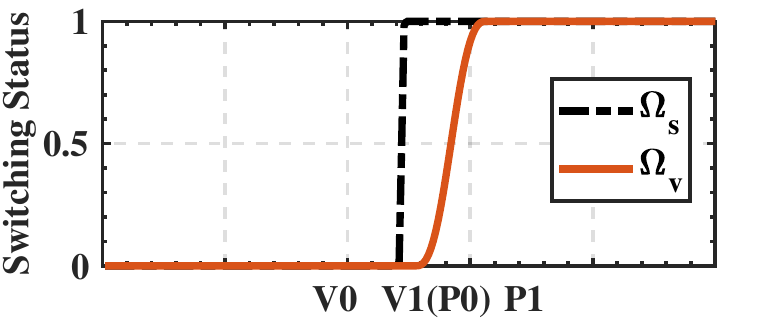}
	\caption{Sketch map of the System Switching Strategy}      
	\label{sketchmap}     
\end{figure}

A sketch map of the transition progress of $\Omega_{s}$ and $\Omega_{v}$ are illustrated in Figure [\ref{sketchmap}].
Accordingly, there exist 4 status such that:

1.$\Omega_{s} = 0,\Omega_{v} = 0$. 

2. $\Omega_{s}= 1, \Omega_{v}\in \left(0,1\right)$. 

3. $\Omega_{s} = 1, \Omega_{v} = 1$. 

4. $\Omega_{s} \in \left(0,1\right), \Omega_{v} = 0$. 

Notably, for case 4, since $\delta$ is a small value, its transition time is short enough that can be practically regarded as 0. On the other hand, the switching of $\Omega_{s}$ only controls the status of the switching PPF, therefore such a short transition progress will have tiny influence on system's stability. In conclusion, we focus on the stability analysis of the system on case $1,2,3$, which stands for the "PPC-only", "PPC and APF" and "APF-only" condition in this paper. These 3 conditions will be the main part of our following analysis.

\section{Control Law Derivation}
In this section, based on the aforementioned system switching strategy in section \ref{SW}, a switching backstepping controller is developed to realize the desired control objective. The prescribed performance function and the backstepping controller's virtual control law will switching accordingly for different condition, and the tracking differentiator will ensure that the system is able to track the calculated virtual control law.

\textbf{Step 1.} Define the first-layer error subsystem as $e_{1} = \varepsilon_{q}$. View the angular velocity as the virtual control law, denoted as $\boldsymbol{v}$, the virtual control law $\boldsymbol{v}$ is presented as follows:
\begin{equation}
	\label{virtual}
	\begin{aligned}
			     \boldsymbol{v} &= \frac{\boldsymbol{r}^{\times}_{b}\boldsymbol{B}_{b}}{\|\boldsymbol{r}^{\times}_{b}\boldsymbol{B}_{b}\|^{2}}\left(-K_{1}\rho_{q}\varepsilon_{q}\right)\left(1-\Omega_{v}\right)\\
			     &\quad -\frac{\left(k_{a}\boldsymbol{r}^{\times}_{b}\boldsymbol{B}_{b} - k_{r}\nabla_{\beta}\boldsymbol{f}^{\times}_{b}\boldsymbol{B}_{b}\right)}{\|\left(k_{a}\boldsymbol{r}^{\times}_{b}\boldsymbol{B}_{b} - k_{r}\nabla_{\beta}\boldsymbol{f}^{\times}_{b}\boldsymbol{B}_{b}\right)\|^{2}}\Omega_{v}K_{p}
	\end{aligned}
\end{equation}
where $K_{1}>0$ denotes the controller gain of the "PPC" part, $K_{p}>0$ represents the gain of the "APF" part.
 Considering the following candidate Lyapunov function for the $e_{1}$ layer, expressed as follows:
\begin{equation}
	   V_{q} = gF\ln\left[\cosh\left(\varepsilon_{q}/F\right)\right] + U
\end{equation}
where $U = U_{a} + U_{r}$ stands for the "APF" potential field.
Take the time-derivative of $V_{q}$ and substituting the result in equation (\ref{deq}), we have:
\begin{equation}
	\label{dVq}
	\begin{aligned}
		\dot{V}_{q} &= g\tanh\left(\frac{\varepsilon_{q}}{F}\right)\frac{1}{\rho_{q}}\left[\left(\boldsymbol{r}^{\times}_{b}\boldsymbol{B}_{b}\right)^{\text{T}}\boldsymbol{\omega}_{s}\right]- \frac{\dot{\rho}_{q}}{\rho_{q}}g\tanh\left(\frac{\varepsilon_{q}}{F}\right)\varepsilon_{q}\\
		&\quad+\left[k_{a}\left(\boldsymbol{r}^{\times}_{b}\boldsymbol{B}_{b}\right)-	k_{r}\nabla_{\beta}\left(\boldsymbol{f}^{\times}_{b}\boldsymbol{B}_{b}\right)\right]^{\text{T}}\boldsymbol{\omega}_{s}
	\end{aligned}
\end{equation}

For further analysis, define an error state variable as $\boldsymbol{e}_{2} = \boldsymbol{\omega}_{s} - \boldsymbol{v}$, which represents the tracking error to the virtual control law.

\textbf{Case 1.} When the boresight vector $\boldsymbol{B}_{b}$ is away from the pointing-forbidden direction such that $\beta\in\left[-1,V_{0}\right]$, we have $\Omega_{s} = 0$, $\Omega_{v} = 0$ and $U_{r} = 0$. Substituting the virtual control law (\ref{virtual}) into (\ref{dVq}), one can be obtained that:
\begin{equation}
	\label{dVV}
	\begin{aligned}
		\dot{V}_{q} &= -\left(K_{1}+\frac{\dot{\rho}_{q}}{\rho_{q}}\right)g\tanh\left(\frac{\varepsilon_{q}}{F}\right)\varepsilon_{q}\\
		&\quad+\frac{g}{\rho_{q}}\tanh\left(\frac{\varepsilon_{q}}{F}\right)\left(\boldsymbol{r}^{\times}_{b}\boldsymbol{B}_{b}\right)^{\text{T}}\boldsymbol{e}_{2}\\
		&\quad+k_{a}\left(-K_{1}\rho_{q}\varepsilon_{q}\right)+k_{a}\left(\boldsymbol{r}^{\times}_{b}\boldsymbol{B}_{b}\right)^{\text{T}}\boldsymbol{e}_{2}
	\end{aligned}
\end{equation}
Define $\boldsymbol{P}=\left[\frac{g}{\rho_{q}}\tanh\left(\varepsilon_{q}/F\right)+k_{a}\right]\left(\boldsymbol{r}^{\times}_{b}\boldsymbol{B}_{b}\right)$.
Since $\varepsilon_{q} = x_{e}/\rho_{q}$, we have $-k_{a}K_{1}\rho_{q}\varepsilon_{q} = -k_{a}K_{1}x_{e}$.
Note $U = U_{a}$ is hold for the current situation, hence the equation (\ref{dVV}) can be rearranged as:
\begin{equation}
	\label{EQ23}
	\begin{aligned}
			   \dot{V}_{q}&=-\left(K_{1}+\frac{\dot{\rho}_{q}}{\rho_{q}}\right)g\tanh\left(\frac{\varepsilon_{q}}{F}\right)\varepsilon_{q}-K_{1}U+\boldsymbol{P}^{\text{T}}\boldsymbol{e}_{2}
	\end{aligned}
\end{equation}
According to section \ref{PPFdesign}, we have $\dot{\rho}_{q}/\rho_{q} = -k_{\rho}\left[1-\rho_{\infty}/\rho_{q}\right]$ for current condition.
 For $\rho_{q} >> \rho_{\infty}$ such that the system is at the converge state, it can be approximated that $\frac{\dot{\rho}_{q}}{\rho_{q}} \approx -k_{\rho}$; for $\rho_{q}\to\rho_{\infty}$, $\frac{\dot{\rho}_{q}}{\rho_{q}} \to 0$ will be obtained. 
 Notably, we have the following property.
  \begin{property}
 	\label{P1}
 	Considering the following function $H\left(x\right) = x\tanh x - \ln\cosh x$, it will satisfies $H\left(x\right) \ge 0$ for $x \ge 0$.
 \end{property}

 Further, applying the conclusion in property \ref{P1}, we have $gF\varepsilon_{q}/F\tanh\left(\varepsilon_{q}/F\right)\ge gF\ln\left(\cosh\left(\varepsilon_{q}/F\right)\right)$. Sort out these results yields:
\begin{equation}
	\begin{aligned}
			 \dot{V}_{q} &< -\left(K_{1}-k_{\rho}\right)gF\ln\cosh\left(\varepsilon_{q}/F\right) \\
			 &\quad- \left(K_{1}-k_{\rho}\right)U+\boldsymbol{P}^{\text{T}}\boldsymbol{e}_{2}
	\end{aligned}
\end{equation}
which can be further written as:
\begin{equation}
		\dot{V}_{q} \le-\left(K_{1}-k_{\rho}\right)V_{q}+\boldsymbol{P}^{\text{T}}\boldsymbol{e}_{2}
\end{equation}

\textbf{Case 2.} According to the switching strategy stated in section \ref{SW}, since $\Omega_{v}$ will switch mildly, there exists a time interval such that $\Omega_{s} = 1$, $\Omega_{v}\in\left(0,1\right)$ is hold. 
In view of this condition, according to the analysis in subsection \ref{PPFdesign}, we have:
\begin{equation}
	\left(gF\ln\left[\cosh\left(\varepsilon_{q}/F\right)\right]\right)^{'} = 0
\end{equation}

Taking the time-derivative of $V_{q}$, one has:
\begin{equation}
	\label{dV2}
			\dot{V}_{q} =\left[k_{a}\left(\boldsymbol{r}^{\times}_{b}\boldsymbol{B}_{b}\right)-	k_{r}\nabla_{\beta}\left(\boldsymbol{f}^{\times}_{b}\boldsymbol{B}_{b}\right)\right]^{\text{T}}\boldsymbol{\omega}_{s}
\end{equation}
	It can be observed that the Lyapunov function is only governed by $U_{a}$ and $U_{r}$ now, which indicates that the "PPC" part is deactivated.
Substituting the virtual control law (\ref{virtual}) into (\ref{dV2}), we have:
\begin{equation}
	\label{case2}
	\begin{aligned}
			\dot{V}_{q} &= -k_{a}K_{1}\rho_{q}\varepsilon_{q} \left(1-\Omega_{v}\right)\\
			&\quad- k_{r}\nabla_{\beta}\left(\boldsymbol{f}^{\times}_{b}\boldsymbol{B}_{b}\right)^{\text{T}}\frac{\boldsymbol{r}^{\times}_{b}\boldsymbol{B}_{b}}{\|\boldsymbol{r}^{\times}_{b}\boldsymbol{B}_{b}\|^{2}}\left(-K_{1}\rho_{q}\varepsilon_{q}\right)\left(1-\Omega_{v}\right)\\
			&\quad+\left(k_{a}\boldsymbol{r}^{\times}_{b}\boldsymbol{B}_{b}-	k_{r}\nabla_{\beta}\boldsymbol{f}^{\times}_{b}\boldsymbol{B}_{b}\right)^{\text{T}}\boldsymbol{e}_{2}-\Omega_{v}K_{p}	\end{aligned}
\end{equation}
Considering the expression stated in equation (\ref{case2}), we have:
\begin{equation}
	\left(\boldsymbol{f}^{\times}_{b}\boldsymbol{B}_{b}\right)^{\text{T}}\boldsymbol{r}^{\times}_{b}\boldsymbol{B}_{b} = \|\boldsymbol{f}^{\times}_{b}\boldsymbol{B}_{b}\|\|\boldsymbol{r}^{\times}_{b}\boldsymbol{B}_{b}\|\cos\psi
\end{equation} 
where $\psi$ stands for the angle between the two cross product results expressed as $\boldsymbol{r}^{\times}_{b}\boldsymbol{B}_{b}$ and $\boldsymbol{f}^{\times}_{b}\boldsymbol{B}_{b}$. 
 Define $\theta_{f} = \text{ang}\left(\boldsymbol{f}_{b},\boldsymbol{B}_{b}\right)$, $\theta_{d} = \text{ang}\left(\boldsymbol{r}_{b},\boldsymbol{B}_{b}\right)$.
Accordingly, from the characteristic of cross product, we have the following conclusion:
\begin{equation}
	\label{eq29}
	\begin{aligned}
		\frac{\left(\boldsymbol{f}^{\times}_{b}\boldsymbol{B}_{b}\right)^{\text{T}}\boldsymbol{r}^{\times}_{b}\boldsymbol{B}_{b} }{\|\boldsymbol{r}^{\times}_{b}\boldsymbol{B}_{b}\|^{2}}\ =&\frac{\|\boldsymbol{f}^{\times}_{b}\boldsymbol{B}_{b}\|}{\|\boldsymbol{r}^{\times}_{b}\boldsymbol{B}_{b}\|}\cos\psi=\frac{\sin\theta_{f}}{\sin\theta_{d}}\cos\psi
	\end{aligned}
\end{equation} 
 Notably, for arbitrary $\theta_{f}$, $\sin\theta_{f} \le 1$ will be always hold. Similarly, $\cos\psi \le 1$ will be always hold. Therefore, one can be obtained that:
\begin{equation}
	\label{eq26}
	\begin{aligned}	&\left(k_{a}-k_{r}\nabla_{\beta}\frac{\sin\theta_{f}}{\sin\theta_{d}}\cos\psi\right)\left(-K_{1}x_{e} \right)\left(1-\Omega_{v}\right)\\
		\le&
		\left(k_{a}-\frac{k_{r}\nabla_{\beta}}{\sin\theta_{d}}\right)\left(-K_{1}x_{e} \right)\left(1-\Omega_{v}\right)\\
	\end{aligned}
\end{equation}
Define $\boldsymbol{P}_{1} = k_{a}\boldsymbol{r}^{\times}_{b}\boldsymbol{B}_{b}-	k_{r}\nabla_{\beta}\boldsymbol{f}^{\times}_{b}\boldsymbol{B}_{b}$ for brevity.
 According to the definition in (\ref{UB}), $k_{r}\nabla_{\beta} \le r$ will be always hold. Substituting equation (\ref{eq26}) into $\dot{V}_{q}$, thus we have:
\begin{equation}
	\begin{aligned}
			\dot{V}_{q} &\le \left(k_{a}-\frac{r}{\min\left(\sin\theta_{d}\right)}\right)\left(-K_{1}x_{e}\right)\left(1-\Omega_{v}\right)\\
			&\quad-\Omega_{v}K_{p}+\boldsymbol{P}^{\text{T}}_{1}\boldsymbol{e}_{2}
	\end{aligned}
\end{equation}
where $\min\left(\sin\theta_{d}\right)$ stands for the minima of $\sin\theta_{d}$ in case 2 condition, and its range will be discussed later in remark \ref{range}.
Specially, let $k_{a}-\frac{r}{\min\left(\sin\theta_{d}\right)} \ge 0$, we have the final conclusion:

\begin{equation}
	\label{conclusion}
	\begin{aligned}
	\dot{V}_{q} & < 
	-\Omega_{v}K_{p}
	+\boldsymbol{P}^{\text{T}}_{1}\boldsymbol{e}_{2}
\end{aligned}
\end{equation}

\textbf{Case 3.} if the system is totally switched to the APF status such that $\Omega_{v} = 1$ and $\Omega_{s} =1$ is hold simultaneously, the system will be governed by APF part of the controller. Taking the time-derivative of $V_{q}$, we have:
\begin{equation}
	\dot{V}_{q} = -K_{p} +\boldsymbol{P}^{\text{T}}_{1}\boldsymbol{e}_{2}
\end{equation}

According to the assumption \ref{ASS2}, the system's target status will be far away from the exclusion zone with a minimum angle $\Theta_{df}$, hence $\rho_{q}$ will finally switch to the "PPF" status and converge to the steady-state.

\begin{remark}
	  Considering the derived virtual control law $\boldsymbol{v}$, it can be mainly divided into two part: the one for "PPC" system governed by $1-\Omega_{s}$ and the one for the "APF" system governed by $\Omega_{s}$.
	  
	  Notably, considering the direction of each part, we can observe that the virtual control law derived from $\boldsymbol{\varepsilon_{q}}$ and the virtual control law derived from the APF's attraction field has the same direction, of which is governed by $\boldsymbol{r}_{b}^{\times}\boldsymbol{B}_{b}$. This indicates us that the effect of the PPC scheme can be regarded as an additional part of the APF's attraction field, and it can be directly added together. This characteristic is owing to the application of the reduced-attitude representation, which stands for the basis of this paper.
\end{remark}

\textbf{Step 2.} In order to ensure that $\boldsymbol{\omega}_{s}$ traces the virtual control law $\boldsymbol{v}$ tightly such that $\boldsymbol{e}_{2}$ is small enough, we employ the Tracking Differentiator (TD) technique to generate the approximation signal for $\dot{\boldsymbol{v}}$. 

Taking the time-derivative of $\boldsymbol{e}_{2}$, we have:
\begin{equation}
	\boldsymbol{J}\dot{e}_{2} =   -\boldsymbol{\omega}^{\times}_{s}\boldsymbol{J}\boldsymbol{\omega}_{s} + \boldsymbol{u} + \boldsymbol{d} - \boldsymbol{J}\dot{v}
\end{equation}
For the tracking differentiator, $\boldsymbol{S}_{d}$ and $\dot{\boldsymbol{S}}_{d}$ stands for the filter output of the tracking differentiator. The actual control law is designed as follows:
\begin{equation}
	\label{ControlLaw}
	\begin{aligned}
			\boldsymbol{u} &= \boldsymbol{\omega}^{\times}_{s}\boldsymbol{J}\boldsymbol{\omega}_{s} - K_{\omega}\boldsymbol{e}_{2} -  D_{m}\text{vec}\left[\tanh\left(\boldsymbol{e}_{2}/\eta\right)\right]+\boldsymbol{J}\dot{\boldsymbol{S}}_{d}\\
			&\quad-g\tanh\left(\varepsilon_{q}/F\right)\frac{1}{\rho_{q}}\boldsymbol{r}^{\times}_{b}\boldsymbol{B}_{b}\left(1-\Omega_{s}\right)\\
			&\quad-\left(k_{a}\boldsymbol{r}^{\times}_{b}\boldsymbol{B}_{b}-	k_{r}\nabla_{\beta}\boldsymbol{f}^{\times}_{b}\boldsymbol{B}_{b}\right)\Omega_{v}
	\end{aligned}
\end{equation}
Considering the following candidate Lyapunov function expressed as $V_{\omega} = \frac{1}{2}\boldsymbol{e}^{\text{T}}_{2}\boldsymbol{J}\boldsymbol{e}_{2}$, take the time-derivative of $\boldsymbol{e}_{2}$, one can be obtained that:
\begin{equation}
	\label{dVo}
	\begin{aligned}
			   \dot{V}_{\omega} &= \boldsymbol{e}^{\text{T}}_{2}\left[-\boldsymbol{\omega}^{\times}_{s}\boldsymbol{J}\boldsymbol{\omega}_{s} + \boldsymbol{u} + \boldsymbol{d} - \boldsymbol{J}\dot{\boldsymbol{v}}\right]\\
			   &=\boldsymbol{e}^{\text{T}}_{2}\left[-K_{\omega}\boldsymbol{e}_{2} + \boldsymbol{d}-D_{m}\text{vec}\left(\tanh\frac{\boldsymbol{e}_{2}}{\eta}\right)\right]\\
			   	&\quad-g\tanh\left(\varepsilon_{q}/F\right)\frac{1}{\rho_{q}}\left(\boldsymbol{r}^{\times}_{b}\boldsymbol{B}_{b}\right)^{\text{T}}\boldsymbol{e}_{2}\left(1-\Omega_{s}\right)\\
			   &\quad-\left[k_{a}\boldsymbol{r}^{\times}_{b}\boldsymbol{B}_{b}-	k_{r}\nabla_{\beta}\boldsymbol{f}^{\times}_{b}\boldsymbol{B}_{b}\right]^{\text{T}}\Omega_{v}\boldsymbol{e}_{2}+\boldsymbol{J}\left[\dot{\boldsymbol{S}}_{d} - \dot{\boldsymbol{v}}\right]\\
	\end{aligned}
\end{equation}
According to the analysis in subsection \ref{TD}, the deviation between $\dot{\boldsymbol{S}}_{d}$ and $\dot{\boldsymbol{v}}$ will be asymptotically converged. In view of this conclusion, we have $\|\dot{\boldsymbol{S}}_{d} - \dot{\boldsymbol{v}}\| \le \epsilon$, where $\epsilon$ is a residual value.
\begin{remark}
	As stated in \ref{TD}, the employed tracking differentiator doesn't relies on the premise that the input signal $\boldsymbol{v}$ is differentiable, which makes it suitable for our control scheme.
\end{remark}
Further, considering the term in equation (\ref{dVo}), according to the mathematical conclusion stated in \cite{walls_globally_2005}, we have:
\begin{equation}
	\begin{aligned}
			&\boldsymbol{e}^{\text{T}}_{2}\boldsymbol{d}-D_{m}\boldsymbol{e}^{\text{T}}_{2}\text{vec}\left(\tanh\frac{\boldsymbol{e}_{2}}{\eta}\right)\\
			\le&\sum_{i=0}^{3}\left(e_{2i}D_{m}-e_{2i}\tanh\frac{e_{2i}}{\eta}D_{m}\right)\\	\le&D_{m}\sum_{i=0}^{3}\left(|e_{2i}|-e_{2i}\tanh\frac{e_{2i}}{\eta}\right)\le 0.8355D_{m}\eta
	\end{aligned}
\end{equation}
Define $D_{0} = 0.8755D_{m}\eta$, this is a constant related to $\eta$. Substituting this result into equation (\ref{dVo}), the expression can be further written as:
\begin{equation}
	\begin{aligned}
		 \dot{V}_{\omega}&\le
		-\boldsymbol{e}^{\text{T}}_{2}K_{\omega}\boldsymbol{e}_{2} + D_{0} + \text{max}\left(\lambda_{J}\right)\epsilon\\
		&\quad-g\tanh\left(\varepsilon_{q}/F\right)\frac{1}{\rho_{q}}\left(\boldsymbol{r}^{\times}_{b}\boldsymbol{B}_{b}\right)^{\text{T}}\boldsymbol{e}_{2}\left(1-\Omega_{s}\right)\\
		&\quad-\left[k_{a}\boldsymbol{r}^{\times}_{b}\boldsymbol{B}_{b}-	k_{r}\nabla_{\beta}\boldsymbol{f}^{\times}_{b}\boldsymbol{B}_{b}\right]^{\text{T}}\Omega_{v}\boldsymbol{e}_{2}
	\end{aligned}
\end{equation}

\textbf{Step 3.}
Define the candidate lumped Lyapunov function $V$ as $V = V_{q} + V_{\omega}$, the analysis of $\dot{V}$ can be divided into three cases.

\textbf{Case 1.} For $\Omega_{s} = 0$, $\Omega_{v} = 0$, the system works under "PPC" mode. Take the time-derivative of $V$, one has:
\begin{equation}
	\begin{aligned}
			\dot{V} &= \dot{V}_{q} + \dot{V}_{\omega}\\ &\le-\left(K_{1}-k_{\rho}\right)V_{q}-2K_{\omega}V_{\omega} + D_{0}+\text{max}\left(\lambda_{J}\right)\epsilon\\
			&\le-\text{min}\left(K_{1}-k_{\rho},2K_{\omega}\right)V+D_{0}+\text{max}\left(\lambda_{J}\right)\epsilon
	\end{aligned}
\end{equation}
where $K_{1}$ and $K_{\omega}$ are all positive controller parameters.

\textbf{Case 2.} For the transition process between "PPC" mode and "APF" mode, the value of $\Omega_{v}$ and $\Omega_{s}$ is $\Omega_{s} = 1$, $\Omega_{v}\in\left(0,1\right)$. Take the time-derivative of $V$, one can be obtained that:
\begin{equation}
	\begin{aligned}
		\dot{V} &\le -\Omega_{v}K_{p}
		-\boldsymbol{e}^{\text{T}}_{2}K_{\omega}\boldsymbol{e}_{2}+D_{0}+\text{max}\left(\lambda_{J}\right)\epsilon
	\end{aligned}
\end{equation}

\textbf{Case 3.} For $\Omega_{s} = 1$, $\Omega_{v} = 1$, the system totally works under "APF" part of the controller, hence we have:
\begin{equation}
	\begin{aligned}
			\dot{V} &\le -K_{p}-\boldsymbol{e}^{\text{T}}_{2}K_{\omega}\boldsymbol{e}_{2}+D_{0}+\text{max}\left(\lambda_{J}\right)\epsilon
	\end{aligned}
\end{equation} 
For each case, $\dot{V} < 0$ can be satisfied. Notably, it can be observed that \textbf{case 1} has the strongest inequality for $\dot{V}$, and the strictly inequality for $\dot{V} < 0$ can be easily obtained in this condition.
 In conclusion, since $D_{0} + \max\left(\lambda_{J}\right)$ can be made small enough for the system, the system will practically asymptotically converge to the steady-state under each condition. 
 
\textbf{Principle of Parameter Selecting}
Additionally, we give some significant principles for the parameter selecting of the proposed controller. 

\textbf{1.} In case 1, it should be guaranteed that $K_{1}-k_{\rho} > 0$ is satisfied. This condition can be well satisfied through parameter selecting.

\textbf{2.} In case 2, note that $k_{a}-\frac{r}{\min\left(\sin\theta_{d}\right)}\ge 0$ should be satisfied, this will govern the selecting of $k_{a}$. 
\begin{remark}
	\label{range}
		Considering the range of $\min\left(\theta_{d}\right)$, 
	owing to the characteristic of sine function, the minima of $\sin\theta_{d}$ for case 2 may be reached at $\min\left(\theta_{d}\right)$ or $\max\left(\theta_{d}\right)$.
	Note that for $\cos\psi = 1$, the vector $\boldsymbol{r}^{\times}_{b}\boldsymbol{B}_{b}$ and $\boldsymbol{f}^{\times}_{b}\boldsymbol{B}_{b}$ will be coplanar.
	Since the system satisfies $\beta\in\left[P_{0},P_{1}\right]$ for case 2, we have:
	\begin{equation}
		\begin{aligned}
			\min\theta_{d} &= \left(\text{ang}\left(\boldsymbol{f}_{b},\boldsymbol{r}_{b}\right)\right) - \arccos P_{0}\\
			\max\theta_{d} &= \left(\text{ang}\left(\boldsymbol{f}_{b},\boldsymbol{r}_{b}\right)\right) - \arccos P_{1}
		\end{aligned}
	\end{equation}
	Notably, according to the assumption \ref{ASS2}, the minimum angle between $\boldsymbol{f}_{b}$ and $\boldsymbol{r}_{b}$ satisfies $\text{ang}\left(\boldsymbol{r}_{b},\boldsymbol{f}_{b}\right) \ge \Theta_{df}$, thus we can yield $\min\theta_{d} \ge \Theta_{df} - \arccos P_{0}$, $\max\theta_{d} \le \pi - \arccos P_{1}$. By compare $\Theta_{df} - \arccos P_{0}$ and $\pi - \arccos P_{1}$, we can have the approximation of $\min\theta_{d}$, which can used for indicating a $k_{a}$.
\end{remark}

\textbf{3.} Notably, there exists two singularity point such that $\|\boldsymbol{r}^{\times}_{b}\boldsymbol{B}_{b}\| = 0$ is hold. This problem can be technically solved by adding a small value $\sigma$ on it. Similarly, another potential singularity point such that $k_{a}\boldsymbol{r}^{\times}_{b}\boldsymbol{B}_{b}-	k_{r}\nabla_{\beta}\boldsymbol{f}^{\times}_{b}\boldsymbol{B}_{b} = \boldsymbol{0}$ can be solved in this way technically.

\textbf{4.} The parameter selecting should ensure that the pointing-forbidden region will not be reached.
Considering the most critical condition such that $\boldsymbol{r}^{\times}_{b}\boldsymbol{B}_{b}$ is collineared with $\boldsymbol{f}^{\times}_{b}\boldsymbol{B}_{b}$, therefore the equilibrium point of the APF is $U_{a} = U_{r}$ such that:
\begin{equation}
	k_{a}x_{E} = k_{r}
\end{equation}

where $x_{E}$ stands for the current value of $x_{e}$ when the boresight vector is on the edge of the exclusion zone such that $\beta=L_{1}$.  This will ensure that the boundary of the exclusion zone is an invariant set of the system.
\section{Simulation Results and Analysis}

In this section, several groups of simulation results and a comparison are presented to show the effectiveness of the proposed scheme.

The spacecraft is supposed to be a rigid-body one with $\boldsymbol{J} =  \text{diag}\left(5.08,5.14,5\right)Kg\cdot m^{2}$.The maximum controller output of this spacecraft is supposed to be $0.5N\cdot m$ for each axis. The external disturbance model is a common one, expressed as follows:
\begin{equation}
	\boldsymbol{d} = 
	\begin{bmatrix}
		1e-3 \cdot \left[4\sin\left(3\omega_{p}t\right) + 3\cos\left(10\omega_{p}t\right) -40\right]\\ 	
		1e-3\left[-1.5\sin\left(2\omega_{p}t\right) + 3\cos\left(5\omega_{p}t\right) +45\right]\\ 	
		1e-3\left[3\sin\left(10\omega_{p}t\right) - 8\cos\left(4\omega_{p}t\right) +40\right]\\ 	
	\end{bmatrix}
\end{equation}
where $\omega_{p} = 0.01$.

\subsection{Obstacle-Avoidance Cases Simulation}
 In this subsection, the spacecraft is required to reorient its boresight axis to a given direction, while the given pointing-forbidden constraint should not be violated. The boresight vector is fixed in the body-fixed frame such that $\boldsymbol{B}_{b} = \left[0,0,1\right]^{\text{T}}$.
 
  The initial attitude and angular velocity of spacecrafts is set to $\boldsymbol{q}_{s}\left(0\right) = \left[0,0,0,1\right]^{\text{T}}$ and $\boldsymbol{\omega}_{s}\left(0\right) = \left[0,0,0\right]^{\text{T}}$ respectively, while the desired pointing vector expressed in the inertial frame is $\boldsymbol{r}_{i} = \left[-0.866,0.5,0\right]^{\text{T}}$.
  
  The performance requirements of this task is established as follows: 1.The system should guarantee that the terminal pointing accuracy should be no more than $0.1^{\circ}$ after $80s$. \textbf{2.} The system should converge to the stable status in no more than $50s$, with a pointing error smaller than $1^{\circ}$.
  According to these requirements, we set the parameter of S-PPF as $\rho_{q}\left(0\right) = 3$, $k_{\rho} = 0.1$, $\rho_{\infty} = 1e-3$. 
  
  \textbf{1. Single Pointing-Forbidden Zone Cases}
  
\begin{figure}[hbt!]
  	\centering 
  	\includegraphics[scale = 0.8]{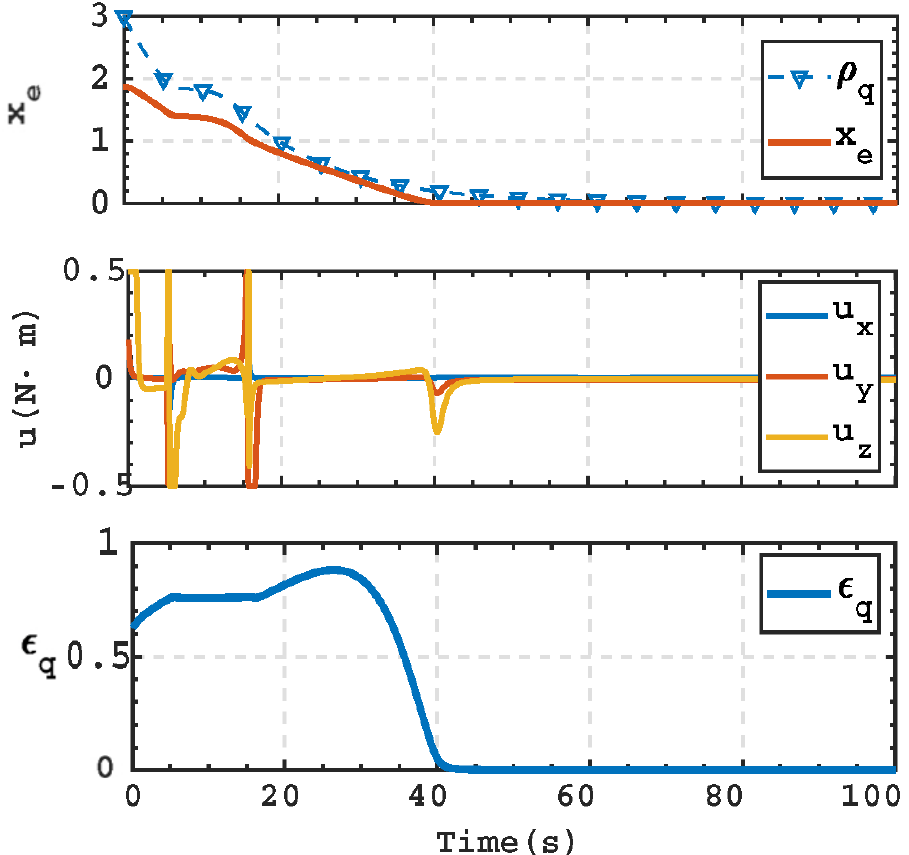}
  	\caption{Time evolution of $x_{e}$, $\boldsymbol{u}$ and $\varepsilon_{q}$(Single Obstacle  $\boldsymbol{f}_{i} = \left[0.5145,0.8575,0\right]^{\text{T}}$)}      
  	\label{singlecons_xe}     
  	\centering 
  	\includegraphics[scale = 0.7]{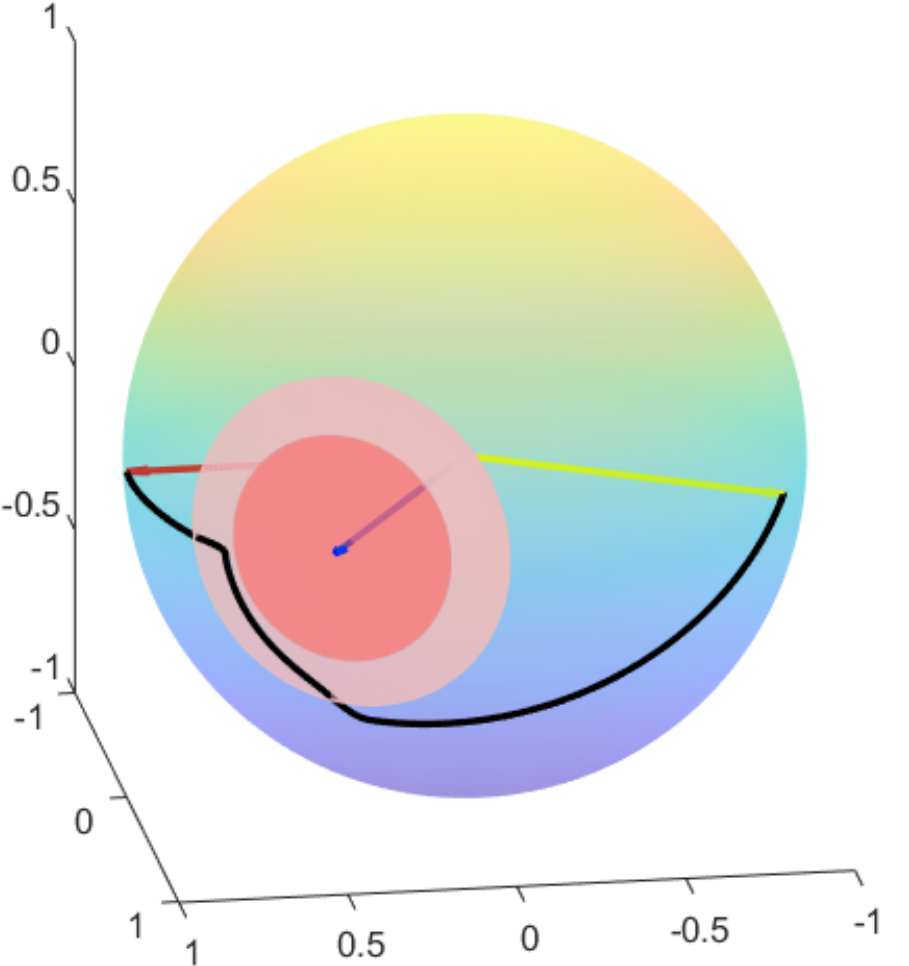}
  	\caption{3-D trajectory of boresight vector $\boldsymbol{B}_{b}$ expressed in the inertial frame(Single Obstacle $\boldsymbol{f}_{i} = \left[0.5145,0.8575,0\right]^{\text{T}}$)}      
  	\label{singlecons_3d1}  
  \end{figure}

  Suppose that the forbidden direction expressed in $\mathfrak{R}_{i}$ is $\boldsymbol{f}_{i} = \left[0.5145,0.8575,0\right]^{\text{T}}$, the simulation results are presented as follows: Figure [\ref{singlecons_xe}] shows the evolution of $x_{e}$, $\boldsymbol{u}$ and $\varepsilon_{q}$ during the whole control procedure, the trajectory of the boresight vector in 3-D sphere is illustrated in Figure [\ref{singlecons_3d1}].
  
  In Figure [\ref{singlecons_xe}], the dotted blue line represents the $\rho_{q}$ trajectory.
  In Figure [\ref{singlecons_3d1}], the black line stands for the trajectory of $\boldsymbol{B}_{i}$ expressed in the $\mathfrak{R}_{i}$ frame. The deep red cone represents the forbidden zone such that $\beta \ge L_{1}$ is hold, while the pink ring area stands for the buffer zone where the repulsion field started working. The red arrow denotes the initial position of the boresight vector, while its desired position is illustrated in yellow.

  From Figure [\ref{singlecons_xe}][\ref{singlecons_3d1}], it can be discovered that the sensor's boresight vector $\boldsymbol{B}_{b}$ successfully circumvents the obstacle zone, and $\varepsilon_{q}$ achieves the desired status at the steady-state in $40s$. From the $3$th subfigure of Figure [\ref{singlecons_xe}] we can discover that there's a time slot at around $6s\to15s$ where the translated variable remain unchanged. Meanwhile, we can observe that the S-PPF is not exponential-converged. This indicates that the Switching-PPF is working under \textbf{Mode 1}, under which the "PPC" is frozen, and will have no impact on the varying of system Lyapunov function.

\begin{figure}[hbt!]   
  	\centering 
  	\includegraphics[scale = 0.6]{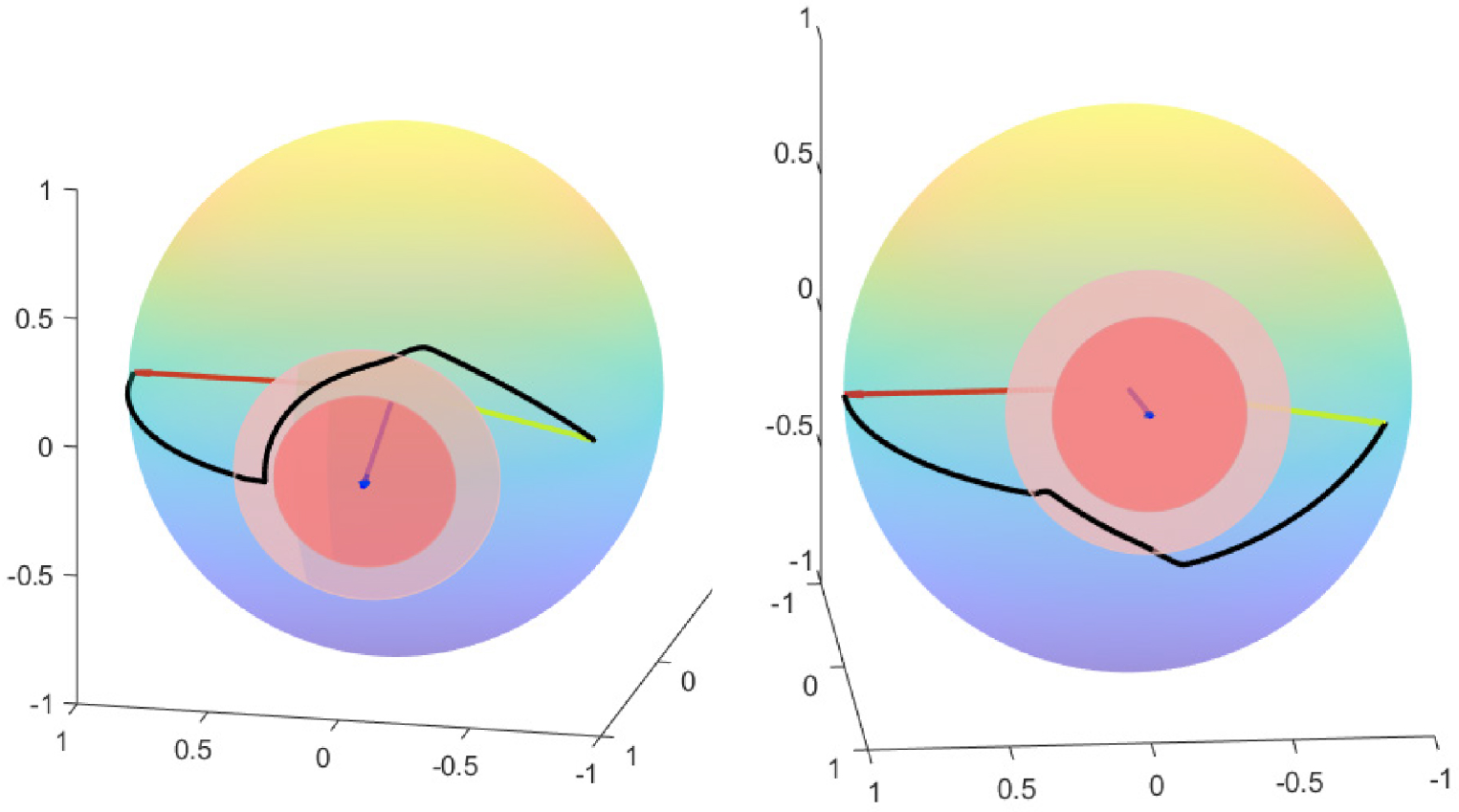}
  	\caption{3-D trajectory of boresight vector $\boldsymbol{B}_{b}$ expressed in the inertial frame(Single Obstacle $\boldsymbol{f}_{i} = \left[-0.099,0.990,-0.099\right]^{\text{T}}$ and $\left[0,0.980,0.196\right]^{\text{T}}$ )}      
  	\label{singleconsim} 
  \end{figure}

  To strength the reliability of the result, another two pointing forbidden zone simulation cases are established for the validation of the proposed scheme. 3-D trajectory is illustrated in Figure [\ref{singleconsim}] respectively, of which the corresponding pointing constraint is set to be $\left[-0.099,0.990,-0.099\right]^{\text{T}}$ and  $\left[0 ,0.980, 0.196\right]^{\text{T}}$.

\begin{figure}[hbt!]
	  	\centering 
	\includegraphics[scale = 0.95]{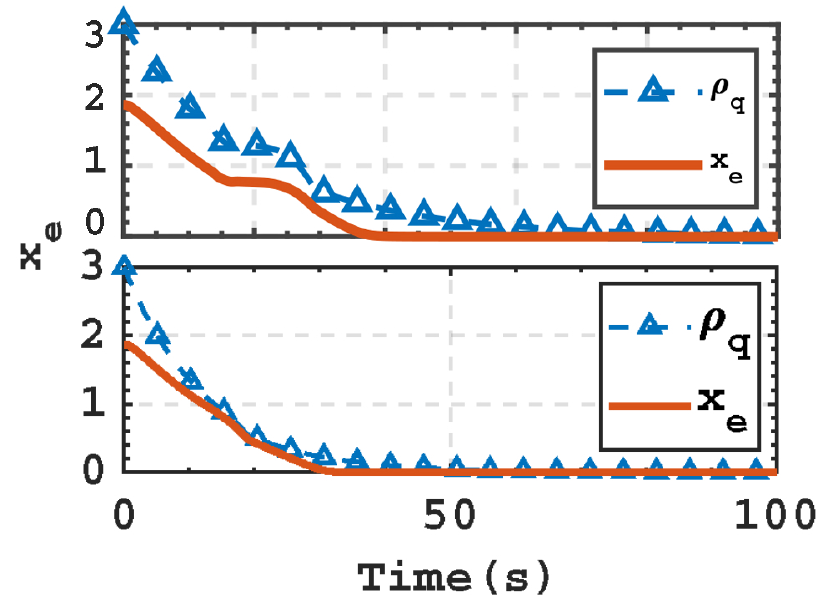}
	\caption{Time evolution of $x_{e}$ (Single Obstacle $\boldsymbol{f}_{i} = \left[-0.099,0.990,-0.099\right]^{\text{T}}$ and $\left[0,0.980,0.196\right]^{\text{T}}$ )}      
	\label{singlecons_XEMIX} 
\end{figure}

 Further, their corresponding $x_{e}$ and its corresponding $\varepsilon_{q}$ trajectories are illustrated in Figure [\ref{singlecons_XEMIX}].
It can be observed that although it takes a bit more time for the simulation case 1 to circumvent the obstacle region, it still achieves the steady-state in the preassigned time. For the simulation case 2, it arrives the desired steady-state region in a short time rapidly.

\begin{figure}[hbt!]
	\centering 
	\includegraphics[scale = 0.95]{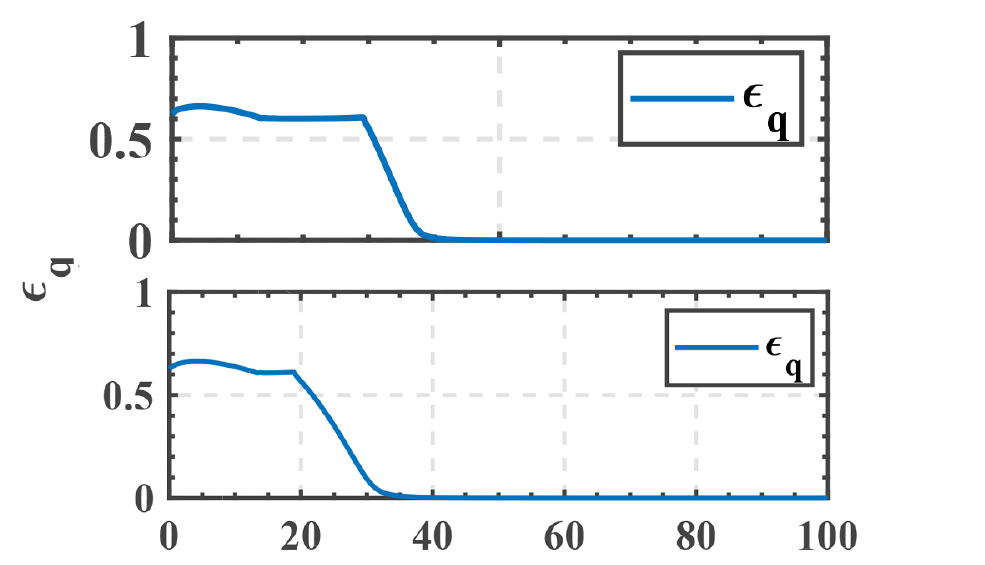}
	\caption{Time evolution of $\varepsilon_{q}$ (Single Obstacle $\boldsymbol{f}_{i} = \left[-0.099,0.990,-0.099\right]^{\text{T}}$ and $\left[0,0.980,0.196\right]^{\text{T}}$ )}      
	\label{epsilon_q_XEMIX} 
\end{figure}

  \textbf{2. Multiple Pointing-Forbidden Zone Cases}
  
  Suppose that there exist multiple pointing-forbidden zones in the $\mathbb{R}^{3}$ space. Considering the forbidden direction listed in Table  [\ref{tab:Settingone}].

\begin{table}[hbt]\centering
	\begin{tabular}{|c|l|c|}
		\hline
		{[}0.571  0.816   0.081{]} & {[}-0.336   0.842   0.421{]} & None  \\ \hline
		{[}0.512   0.854  0.085{]} & {[}-0.188  0.940  -0.282{]} & None          \\ \hline
		{[} 0.514   0.857  0{]} & {[}-0.311   0.778  -0.544{]} & None          \\ \hline
		{[}0.472  0.788 0.394{]} & {[}-0.369   0.924  -0.092{]} & None        \\ \hline
		{[}0.472   0.788 0.394{]} & {[} -0.336   0.842 0.421{]} & {{[} 0.169   0.845  -0.507{]}} \\ \hline
	\end{tabular}
	\caption{\label{tab:Settingone} Simulation vector setting (2 Obstacles and 3 Obstacles)}
\end{table}

\begin{figure}[hbt!]
	\centering 
	\includegraphics[scale = 0.8]{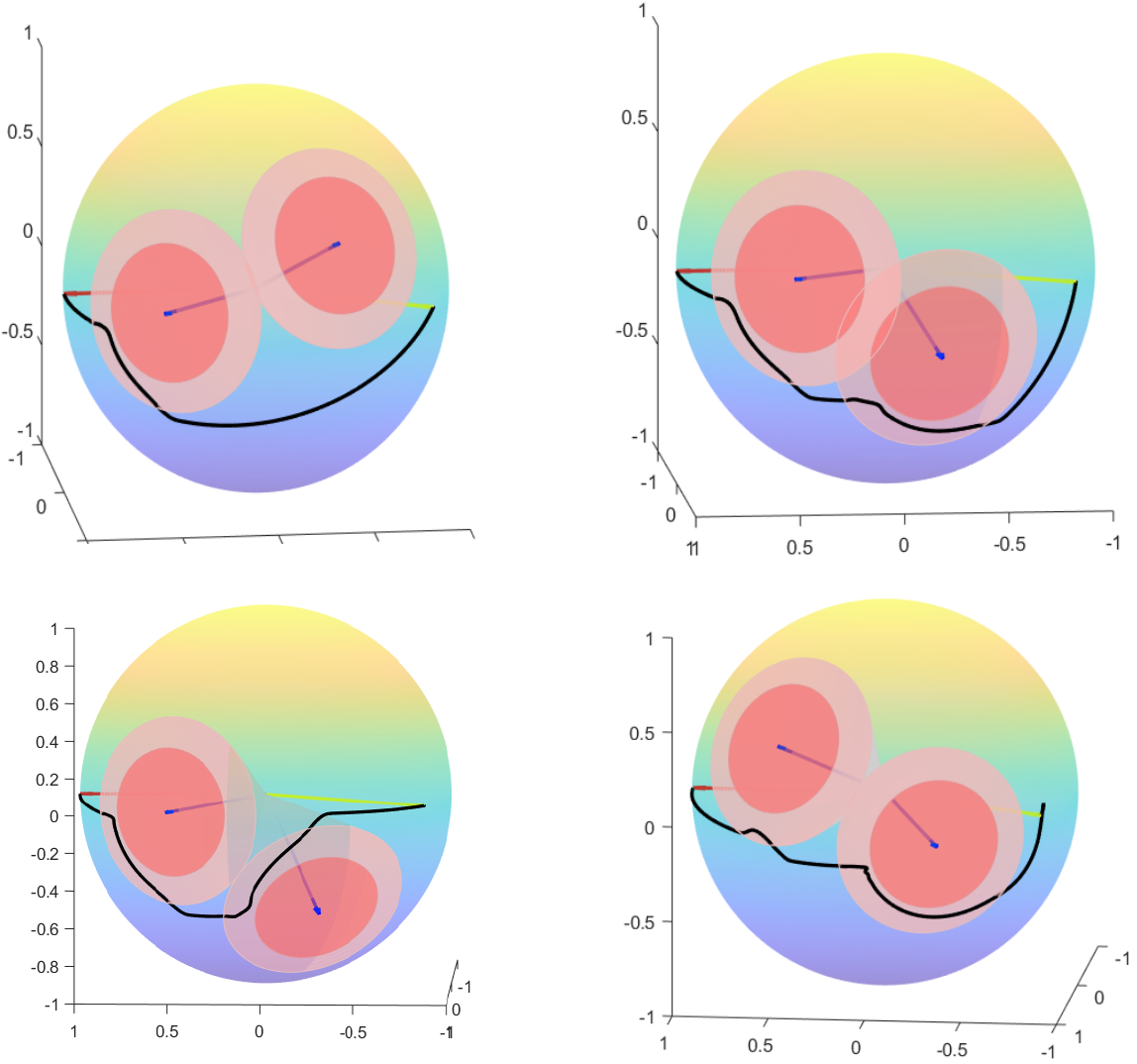}
	\caption{3-D trajectory of boresight vector $\boldsymbol{B}_{b}$ expressed in the inertial frame(2 Obstacles, corresponding to table [\ref{tab:Settingone}])}      
	\label{twoconssim}    
\end{figure}

\begin{figure}[hbt!] 
	\centering 
	\includegraphics[scale = 0.65]{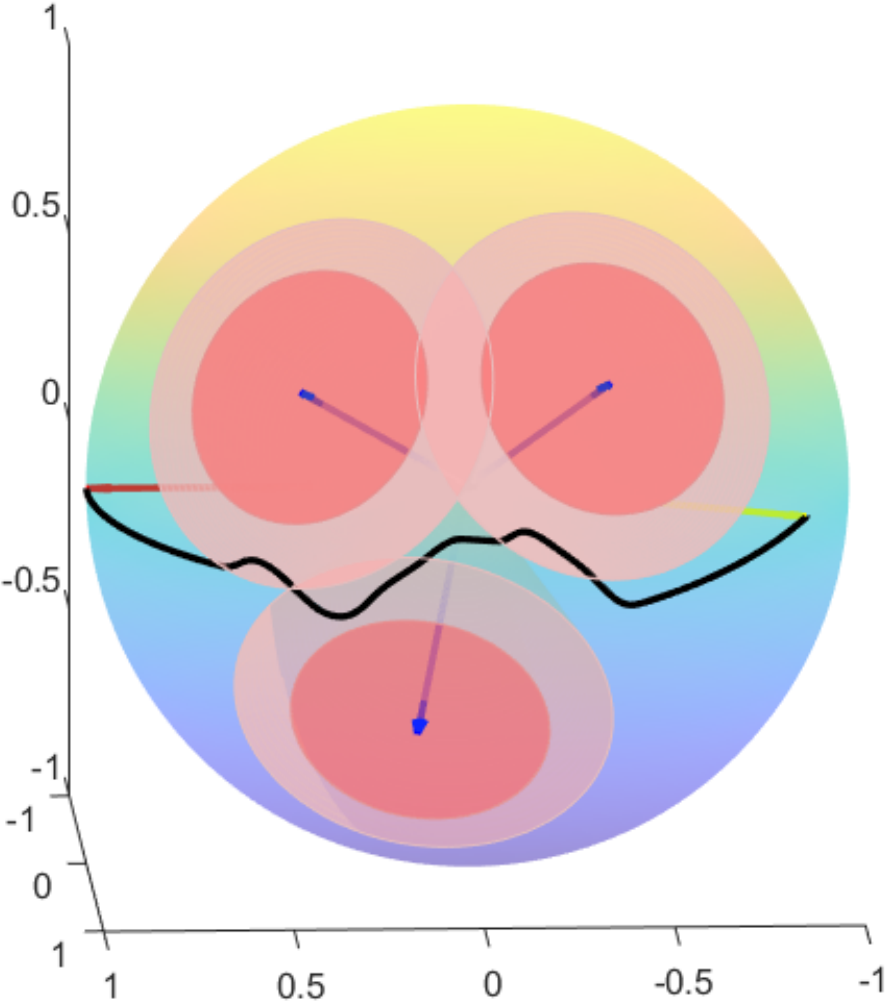}
	\caption{3-D trajectory of boresight vector $\boldsymbol{B}_{b}$ expressed in the inertial frame(3 Obstacles)}      
	\label{THREEconssim}   
\end{figure}

The 3-D trajectory of the boresight vector $\boldsymbol{B}_{b}$ in attitude sphere is shown in Figure [\ref{twoconssim}] and [\ref{THREEconssim}] respectively, where the forbidden zones are illustrated in red colors, same as stated before.


We can observe that the $\boldsymbol{B}_{b}$ trajectory is able to circumvent the pointing-forbidden zone in every case. For case 2, since the two exclusion zone intersects heavily, therefore it is not economic enough to cross through two obstacle. We can find that the trajectory is leading to another direction, bypassing the obstacle and finally reach the desired point.

\begin{figure}[hbt!]
	\centering 
	\includegraphics[scale = 0.6]{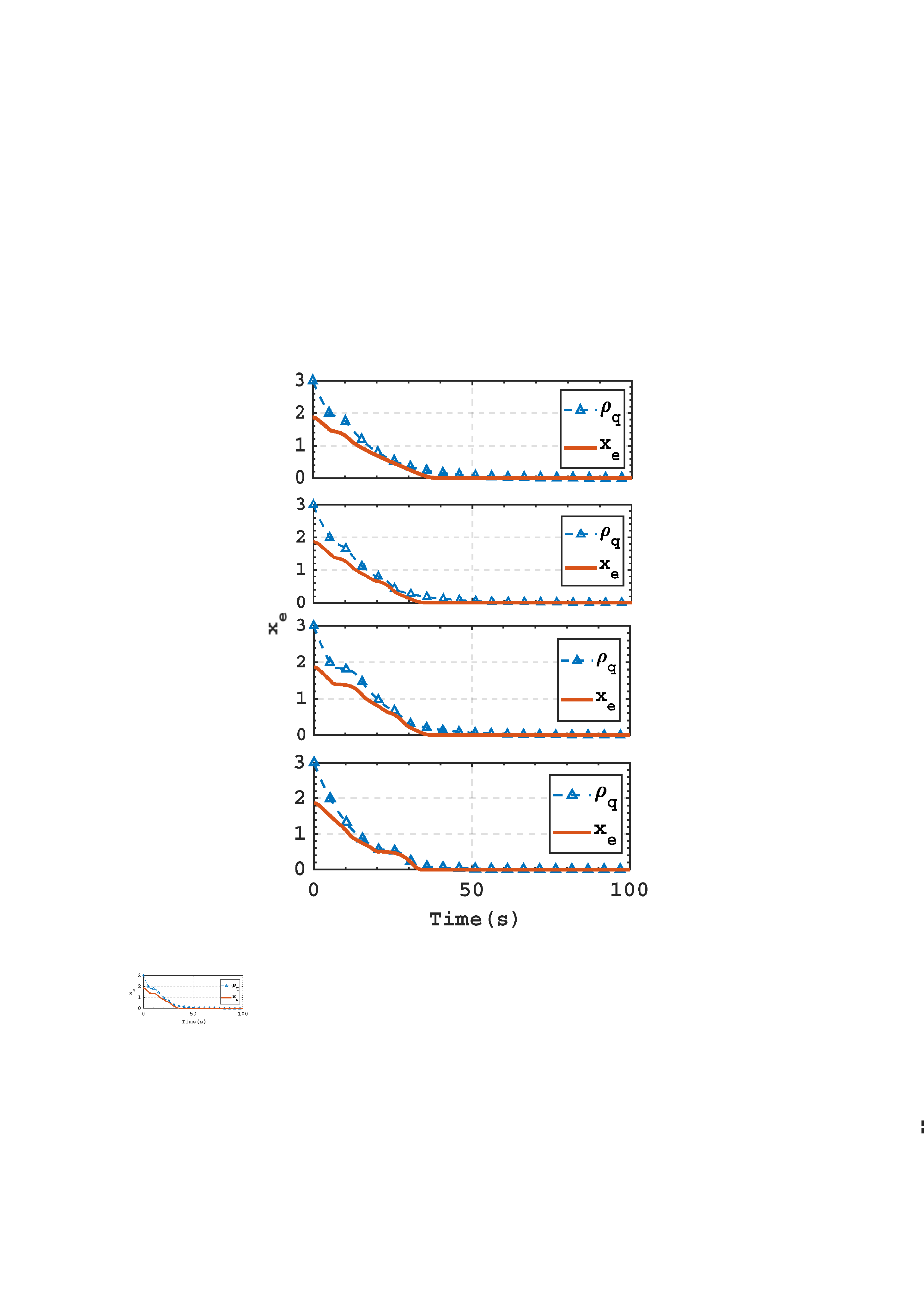}
	\caption{Time evolution of $x_{e}$ trajectory (2 Obstacles, corresponding to Table [\ref{tab:Settingone}])}      
	\label{twocons_xe}    
\end{figure}

\begin{figure}[hbt!]
	\centering 
	\includegraphics[scale = 0.6]{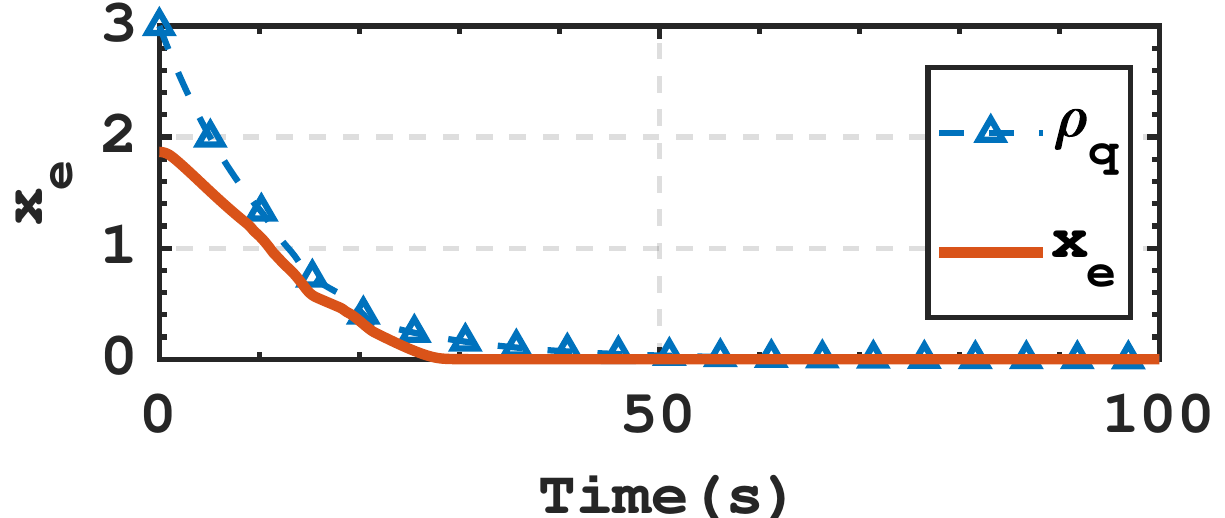}
	\caption{Time evolution of $x_{e}$ trajectory (3 Obstacles)}      
	\label{threecons_xe}  
\end{figure}

Additionally, the $x_{e}$ trajectory of these cases are illustrated in Figure [\ref{twocons_xe}][\ref{threecons_xe}]. We can observe that the system is able to reach the steady-state rapidly in each case.

\subsection{Comparison to APF-only Algorithm}

In this section, a comparison to the traditional APF-only controller is performed to show the effectiveness of the proposed scheme. To establish a comparison, we introduce the APF controller stated in \cite{dongare2021attitude} into consideration.

\begin{figure}[hbt!] 
	\centering 
	\includegraphics[scale = 0.6]{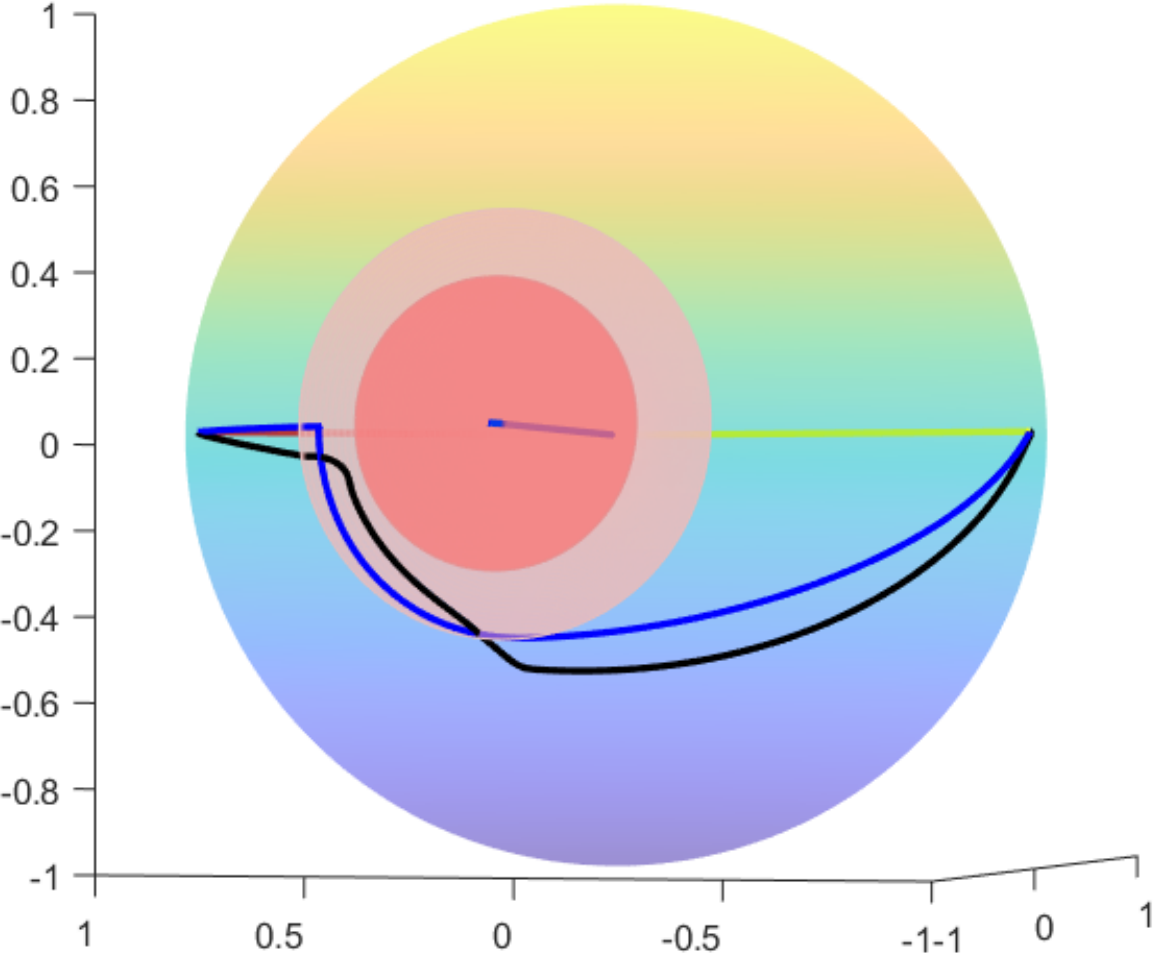}
	\caption{3-D trajectory of boresight vector $\boldsymbol{B}_{b}$ expressed in the inertial frame(Comparison of Benchmark controller 1 and the proposed scheme)}      
	\label{compare_3d}     
	\centering 
	\includegraphics[scale = 0.5]{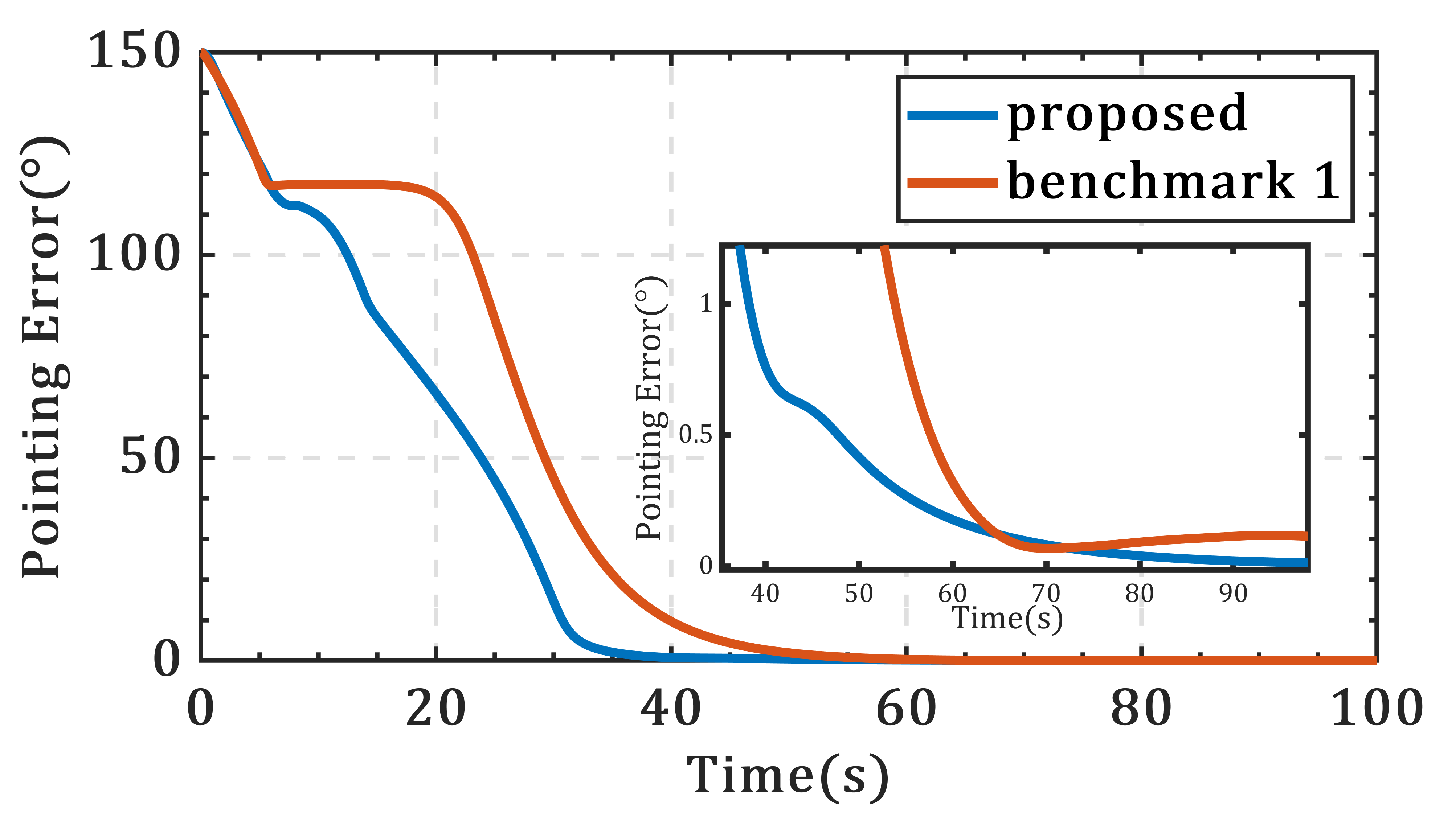}
	\caption{Pointing Error to the target position(Comparison of Benchmark controller 1 and the proposed scheme)}      
	\label{compare_xe}   
\end{figure}

The APF-only controller is named as "benchmark controller 1" for brevity. Considering the pointing-forbidden vector $\boldsymbol{f}_{i} = \left[0.5145,0.8575,0\right]^{\text{T}}$, Figure [\ref{compare_xe}] shows the time responding of the pointing error under each controller, while its 3-D trajectory of $\boldsymbol{B}_{b}$ expressed in inertial frame is illustrated in Figure [\ref{compare_3d}]. The black trajectory denotes the result of the proposed scheme, while the blue one stands for the benchmark controller 1.
\begin{remark}
	Since the proposed control scheme is built based on the \textbf{Backstepping} methodology, the desired control can be only guaranteed when $\boldsymbol{\omega}_{s}$ tracks the virtual control law $\boldsymbol{v}$ tightly.
\end{remark}

For the proposed scheme, it is able to converge to the steady-state in about 40s with a smooth transient behaviour, while the settling time for benchmark controller 1 is longer, as illustrated in Figure [\ref{compare_xe}]. Also, we can find that the terminal control error of the proposed scheme is $0.07^{\circ}$, which is much more smaller than the benchmark controller 1. This validates the necessity of the combination of APF and PPC.

Further, we take a look at the steady-state, it can be discovered that the proposed scheme has a better robustness against the external disturbance, with a better control accuracy is achieved. This is also owing to the application of PPC scheme, as the PPF boundary will exert appropriate repulsion to the state trajectory, forcing it closer to the desired position.

\section{Conclusion}
This paper focuses on the attitude reorientation control problem under pointing-forbidden constraint and the performance constraint. Based on the reduced-attitude representation, we directly fuse the prescribed performance control(PPC) scheme and the artificial potential field(APF) scheme together, and a switching controller structure is proposed. Further, to tackle the intrinsic contradiction between PPC and APF when meets obstacles, a special performance function is presented, ensuring the system's stability. To alleviate the impact caused by the status switching, a particular function is designed to be the mollification of the standard $0-1$ switching function, and the transition process can be adjusted through parameter regulating explicitly. Following these ideas, a complete backstepping controller is presented in this paper, and numerical simulation results have shown its effectiveness for handling different scenarios. This paper presents a possibility of handling the performance constraints and pointing constraints without planning, which is of high application value for real control scenarios. Further investigation may be focused on its incorporation with other typical physical limitations.

\bibliographystyle{IEEEtran}
\bibliography{pointingreference}

\begin{thebibliography}{10}
\providecommand{\url}[1]{#1}
\csname url@samestyle\endcsname
\providecommand{\newblock}{\relax}
\providecommand{\bibinfo}[2]{#2}
\providecommand{\BIBentrySTDinterwordspacing}{\spaceskip=0pt\relax}
\providecommand{\BIBentryALTinterwordstretchfactor}{4}
\providecommand{\BIBentryALTinterwordspacing}{\spaceskip=\fontdimen2\font plus
\BIBentryALTinterwordstretchfactor\fontdimen3\font minus
  \fontdimen4\font\relax}
\providecommand{\BIBforeignlanguage}[2]{{%
\expandafter\ifx\csname l@#1\endcsname\relax
\typeout{** WARNING: IEEEtran.bst: No hyphenation pattern has been}%
\typeout{** loaded for the language `#1'. Using the pattern for}%
\typeout{** the default language instead.}%
\else
\language=\csname l@#1\endcsname
\fi
#2}}
\providecommand{\BIBdecl}{\relax}
\BIBdecl

\bibitem{biggs2016geometric}
J.~D. Biggs and L.~Colley, ``Geometric attitude motion planning for spacecraft
  with pointing and actuator constraints,'' \emph{Journal of Guidance, Control,
  and Dynamics}, vol.~39, no.~7, pp. 1672--1677, 2016.

\bibitem{he2022pointing}
\BIBentryALTinterwordspacing
H.~He, P.~Shi, and Y.~Zhao, ``A pointing-based method for spacecraft attitude
  maneuver path planning under time-varying pointing constraints,''
  \emph{Advances in Space Research}, 2022. [Online]. Available:
  \url{https://doi.org/10.1016/j.asr.2022.05.058}
\BIBentrySTDinterwordspacing

\bibitem{xu2018rotational}
\BIBentryALTinterwordspacing
R.~Xu, H.~Wang, W.~Xu, P.~Cui, and S.~Zhu, ``Rotational-path decomposition
  based recursive planning for spacecraft attitude reorientation,'' \emph{Acta
  Astronautica}, vol. 143, pp. 212--220, 2018. [Online]. Available:
  \url{https://doi.org/10.1016/j.actaastro.2017.11.035}
\BIBentrySTDinterwordspacing

\bibitem{wu2017time}
\BIBentryALTinterwordspacing
C.~Wu, R.~Xu, S.~Zhu, and P.~Cui, ``Time-optimal spacecraft attitude maneuver
  path planning under boundary and pointing constraints,'' \emph{Acta
  Astronautica}, vol. 137, pp. 128--137, 2017. [Online]. Available:
  \url{https://doi.org/10.1016/j.actaastro.2017.04.004}
\BIBentrySTDinterwordspacing

\bibitem{wu2019energy}
C.~Wu and X.~Han, ``Energy-optimal spacecraft attitude maneuver path-planning
  under complex constraints,'' \emph{Acta Astronautica}, vol. 157, pp.
  415--424, 2019.

\bibitem{xu2017rapid}
\BIBentryALTinterwordspacing
R.~Xu, C.~Wu, S.~Zhu, B.~Fang, W.~Wang, L.~Xu, and W.~He, ``A rapid maneuver
  path planning method with complex sensor pointing constraints in the attitude
  space,'' \emph{Information Systems Frontiers}, vol.~19, no.~4, pp. 945--953,
  2017. [Online]. Available: \url{https://doi.org/10.1007/s10796-016-9642-1}
\BIBentrySTDinterwordspacing

\bibitem{hu2019anti}
\BIBentryALTinterwordspacing
Q.~Hu, B.~Chi, and M.~R. Akella, ``Anti-unwinding attitude control of
  spacecraft with forbidden pointing constraints,'' \emph{Journal of Guidance,
  Control, and Dynamics}, vol.~42, no.~4, pp. 822--835, 2019. [Online].
  Available: \url{https://doi.org/10.2514/1.G003606}
\BIBentrySTDinterwordspacing

\bibitem{chi2018reduced}
\BIBentryALTinterwordspacing
B.~Chi, Q.~Hu, and L.~Guo, ``Reduced attitude control in the presence of
  pointing constraint,'' in \emph{2018 37th Chinese Control Conference (CCC)},
  2018, pp. 9781--9785. [Online]. Available:
  \url{https://doi.org/10.23919/ChiCC.2018.8482587}
\BIBentrySTDinterwordspacing

\bibitem{feng2019reorientation}
\BIBentryALTinterwordspacing
Z.-x. Feng, J.-g. Guo, and J.~Zhou, ``Reorientation control for a
  microsatellite with pointing and angular velocity constraints,'' in
  \emph{International Conference on Intelligent Robotics and Applications},
  2019, pp. 717--726. [Online]. Available:
  \url{https://doi.org/10.1007/978-3-030-27532-7_62}
\BIBentrySTDinterwordspacing

\bibitem{bai2021torque}
\BIBentryALTinterwordspacing
Z.~Bai, Y.~Liu, and Q.~Hu, ``Torque-limited attitude control for rigid
  spacecraft with motion constraints,'' in \emph{2021 40th Chinese Control
  Conference (CCC)}, 2021, pp. 7724--7729. [Online]. Available:
  \url{https://doi.org/10.23919/CCC52363.2021.9550402}
\BIBentrySTDinterwordspacing

\bibitem{dongare2021attitude}
\BIBentryALTinterwordspacing
A.~Dongare, R.~Hamrah, and A.~K. Sanyal, ``Attitude pointing control using
  artificial potentials with control input constraints,'' in \emph{2021
  American Control Conference (ACC)}, 2021, pp. 1--6. [Online]. Available:
  \url{https://doi.org/10.23919/ACC50511.2021.9483350}
\BIBentrySTDinterwordspacing

\bibitem{bechlioulis_adaptive_2009}
\BIBentryALTinterwordspacing
C.~P. Bechlioulis and G.~A. Rovithakis, ``Adaptive control with guaranteed
  transient and steady state tracking error bounds for strict feedback
  systems,'' \emph{Automatica}, vol.~45, no.~2, pp. 532--538, 2009. [Online].
  Available: \url{https://doi.org/10.1016/j.automatica.2008.08.012}
\BIBentrySTDinterwordspacing

\bibitem{wei2021overview}
\BIBentryALTinterwordspacing
C.~Wei, Q.~Chen, J.~Liu, Z.~Yin, and J.~Luo, ``An overview of prescribed
  performance control and its application to spacecraft attitude system,''
  \emph{Proceedings of the Institution of Mechanical Engineers, Part I: Journal
  of Systems and Control Engineering}, vol. 235, no.~4, pp. 435--447, 2021.
  [Online]. Available: \url{https://doi.org/10.1177/0959651820952552}
\BIBentrySTDinterwordspacing

\bibitem{shao2018fault}
\BIBentryALTinterwordspacing
X.~Shao, Q.~Hu, Y.~Shi, and B.~Jiang, ``Fault-tolerant prescribed performance
  attitude tracking control for spacecraft under input saturation,'' \emph{IEEE
  Transactions on Control Systems Technology}, vol.~28, no.~2, pp. 574--582,
  2018. [Online]. Available: \url{https://doi.org/10.1109/TCST.2018.2875426}
\BIBentrySTDinterwordspacing

\bibitem{hu2018adaptive}
\BIBentryALTinterwordspacing
Q.~Hu, Y.~Shi, and X.~Shao, ``Adaptive fault-tolerant attitude control for
  satellite reorientation under input saturation,'' \emph{Aerospace Science and
  Technology}, vol.~78, pp. 171--182, 2018. [Online]. Available:
  \url{https://doi.org/10.1016/j.ast.2018.04.015}
\BIBentrySTDinterwordspacing

\bibitem{wei2018learning}
\BIBentryALTinterwordspacing
C.~Wei, J.~Luo, H.~Dai, and G.~Duan, ``Learning-based adaptive attitude control
  of spacecraft formation with guaranteed prescribed performance,'' \emph{IEEE
  transactions on cybernetics}, vol.~49, no.~11, pp. 4004--4016, 2018.
  [Online]. Available: \url{https://doi.org/10.1109/TCYB.2018.2857400}
\BIBentrySTDinterwordspacing

\bibitem{amrr2021prescribed}
\BIBentryALTinterwordspacing
S.~M. Amrr, A.~Alturki, A.~Kumar, and M.~Nabi, ``Prescribed performance-based
  event-driven fault-tolerant robust attitude control of spacecraft under
  restricted communication,'' \emph{Electronics}, vol.~10, no.~14, p. 1709,
  2021. [Online]. Available: \url{https://doi.org/10.3390/electronics10141709}
\BIBentrySTDinterwordspacing

\bibitem{luo2018low}
\BIBentryALTinterwordspacing
J.~Luo, Z.~Yin, C.~Wei, and J.~Yuan, ``Low-complexity prescribed performance
  control for spacecraft attitude stabilization and tracking,'' \emph{Aerospace
  Science and Technology}, vol.~74, pp. 173--183, 2018. [Online]. Available:
  \url{https://doi.org/10.1016/j.ast.2018.01.002}
\BIBentrySTDinterwordspacing

\bibitem{zhang2019prescribed}
\BIBentryALTinterwordspacing
C.~Zhang, G.~Ma, Y.~Sun, and C.~Li, ``Prescribed performance adaptive attitude
  tracking control for flexible spacecraft with active vibration suppression,''
  \emph{Nonlinear Dynamics}, vol.~96, no.~3, pp. 1909--1926, 2019. [Online].
  Available: \url{https://doi.org/10.1007/s11071-019-04894-x}
\BIBentrySTDinterwordspacing

\bibitem{yong2020flexible}
\BIBentryALTinterwordspacing
K.~Yong, M.~Chen, Y.~Shi, and Q.~Wu, ``Flexible performance-based robust
  control for a class of nonlinear systems with input saturation,''
  \emph{Automatica}, vol. 122, 2020. [Online]. Available:
  \url{https://doi.org/10.1016/j.automatica.2020.109268}
\BIBentrySTDinterwordspacing

\bibitem{WANG2022}
\BIBentryALTinterwordspacing
K.~Wang, T.~Meng, W.~Wang, R.~Song, and Z.~Jin, ``Finite-time extended state
  observer based prescribed performance fault tolerance control for spacecraft
  proximity operations,'' \emph{Advances In Space Research}, 2022. [Online].
  Available: \url{https://doi.org/10.1016/j.asr.2022.05.072}
\BIBentrySTDinterwordspacing

\bibitem{guo2011spacecraft}
Y.~Guo, C.~Li, and G.~Ma, ``Spacecraft autonomous attitude maneuver control by
  potential function method,'' \emph{Acta Aeronautica et Astronautica Sinica},
  vol.~32, no.~3, pp. 457--464, 2011.

\bibitem{yang2019active}
\BIBentryALTinterwordspacing
Z.~Yang, J.~Ji, X.~Sun, H.~Zhu, and Q.~Zhao, ``Active disturbance rejection
  control for bearingless induction motor based on hyperbolic tangent tracking
  differentiator,'' \emph{IEEE Journal of Emerging and Selected Topics in Power
  Electronics}, vol.~8, no.~3, pp. 2623--2633, 2019. [Online]. Available:
  \url{https://doi.org/10.1109/JESTPE.2019.2923793}
\BIBentrySTDinterwordspacing

\bibitem{walls_globally_2005}
\BIBentryALTinterwordspacing
R.~J. Wallsgrove and M.~R. Akella, ``Globally stabilizing saturated attitude
  control in the presence of bounded unknown disturbances,'' \emph{J. Guid.
  Control. Dynam}, vol.~28, no.~5, pp. 957--963, 2005. [Online]. Available:
  \url{https://doi.org/10.2514/1.9980}
\BIBentrySTDinterwordspacing

\end{thebibliography}


\begin{IEEEbiography}
	[{\includegraphics[width=1in,height=1.25in,clip,keepaspectratio]{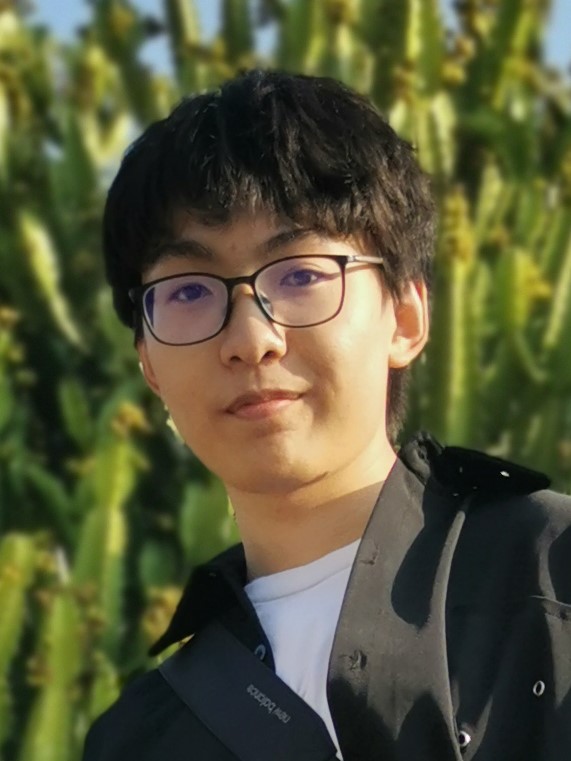}}]
	{Jiakun Lei}{\space} 
	received the B.S. degree from Dep.Automation, University of Electronic Science and Technology of China, Chengdu, China, in 2019. He is currently working toward the Ph.D. degree in aeronautical and astronautical science and technology in Zhejiang University, Hangzhou, China.
	His research interests include attitude planning and control under constraints, attitude control of spacecraft with complex structure and attitude formation.
\end{IEEEbiography}

\begin{IEEEbiography}
	[{\includegraphics[width=1in,height=1.25in,clip,keepaspectratio]{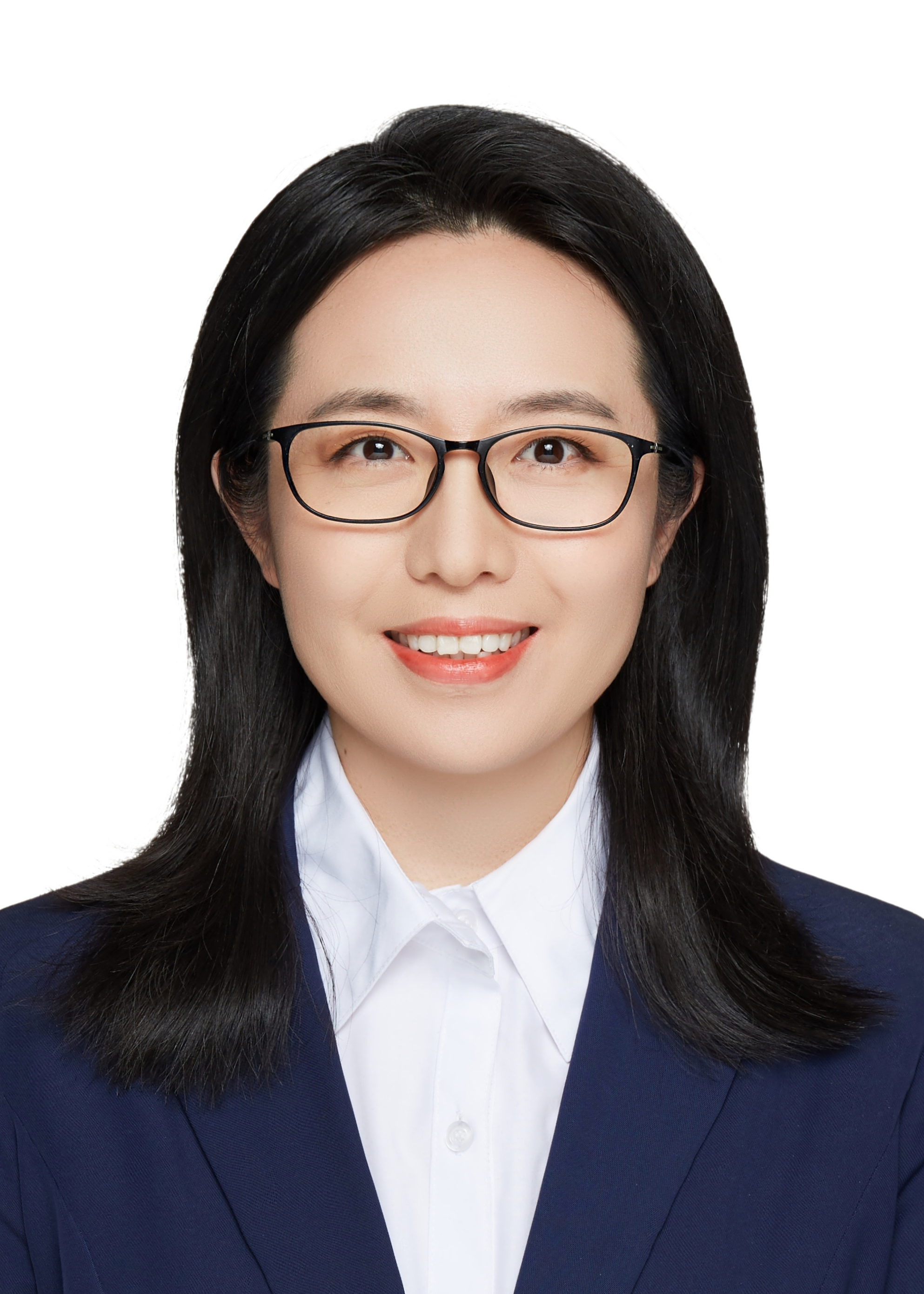}}]
	{Tao Meng}{\space}
	received the B.S. degree from Electronic science and technology, Zhejiang university, Hangzhou, China, in 2004, the M.S. degree from Electronic science and technology, Zhejiang university, Hangzhou, China, in 2006, and the Ph.D. degree from Electronic science and technology, Zhejiang university, Hangzhou, China, in 2009. She is currently the Professor of the School of Aeronautics and astronautics. Her research interest include attitude control, orbit control and constellation formation control of micro-satellite.
\end{IEEEbiography}

\begin{IEEEbiography}
	[{\includegraphics[width=1in,height=1.25in,clip,keepaspectratio]{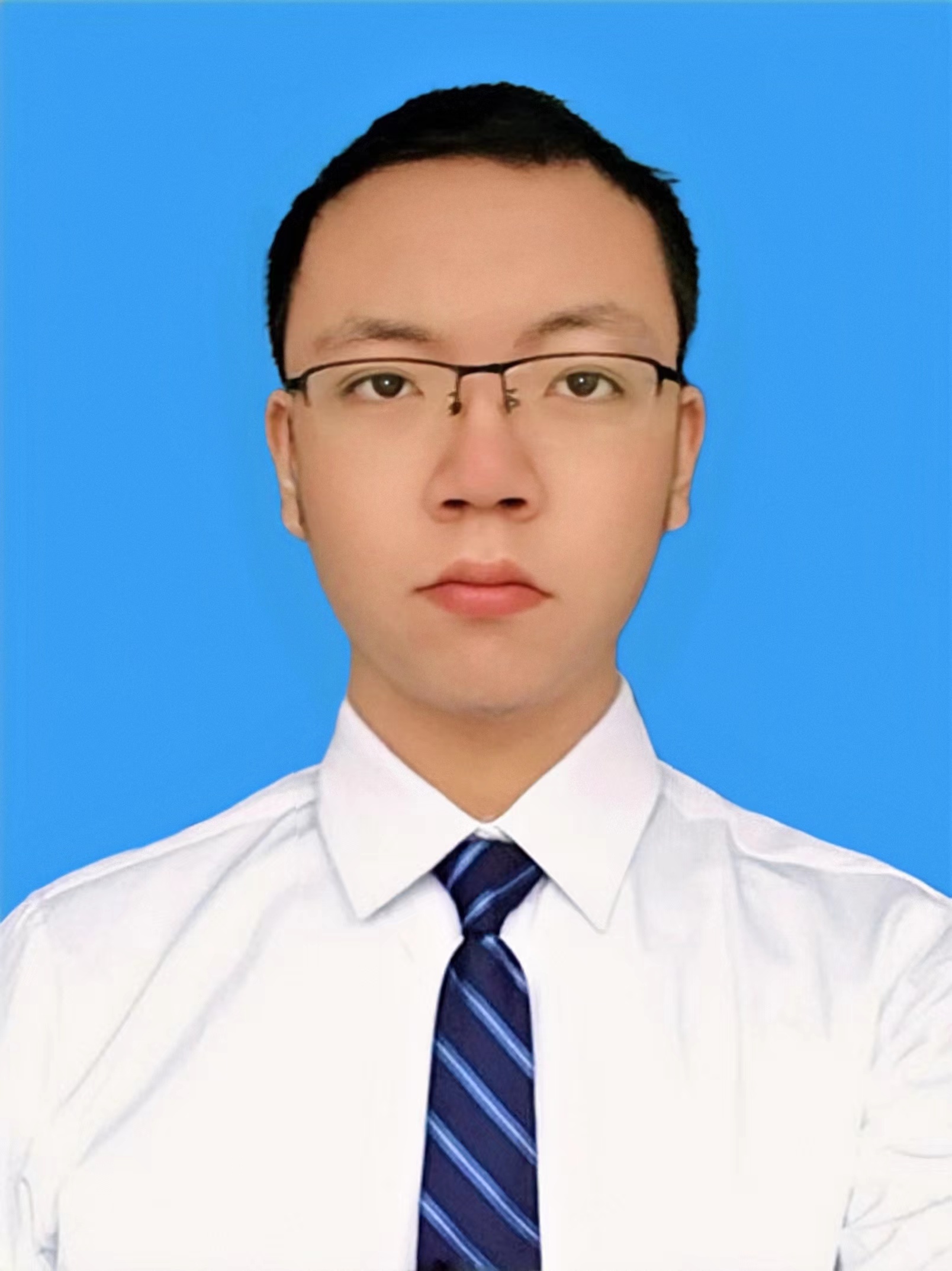}}]
	{Weijia Wang}{\space}
	received the B.S. degree from aerospace engineering, University of Electronic Science and Technology of China, Chengdu, China, in 2020. He is currently working toward the Ph.D. degree in aeronautical and astronautical science and technology in Zhejiang University, Hangzhou, China.	His research interests include model predictive control and reinforcement-learning-based control for 6-DOF spacecraft formation. 
\end{IEEEbiography}

\begin{IEEEbiography}
	[{\includegraphics[width=1in,height=1.25in,clip,keepaspectratio]{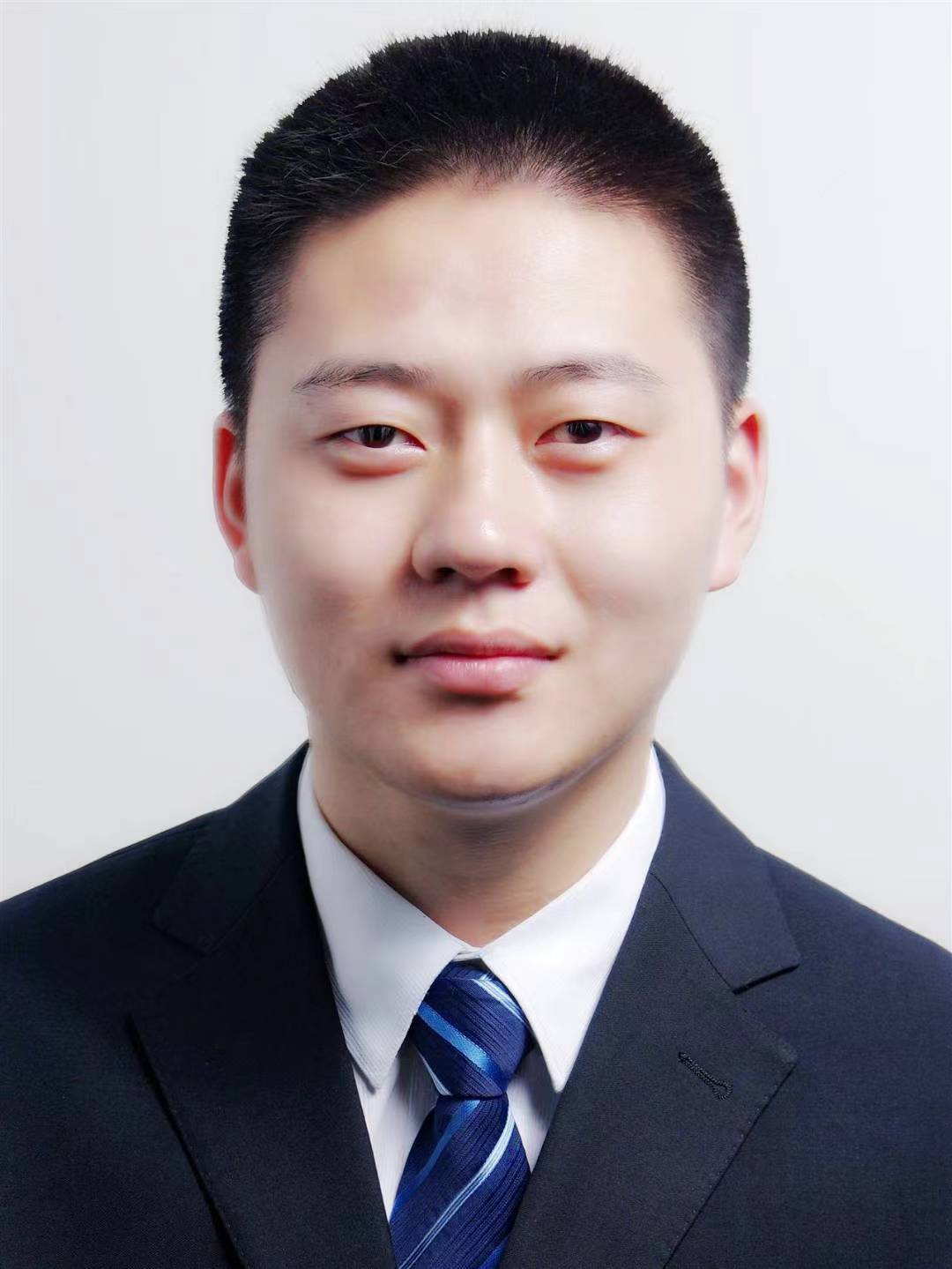}}]
	{Shujian Sun}{\space}
received the B.S. degree from Electronic Information Engineering (Underwater Acoustic), Harbin Engineering University, Harbin, China, in 2013, and the Ph.D. degree from Electronic Science and Technology,Zhejiang University, Hangzhou, China, in 2020. He is recently the Assistant Professor of School of Aeronautics and Astronautics of Zhejiang University. His research interest include orbit control and formation flying of micro-satellite and micro-propulsion technology.
\end{IEEEbiography}

\begin{IEEEbiography}
	[{\includegraphics[width=1in,height=1.25in,clip,keepaspectratio]{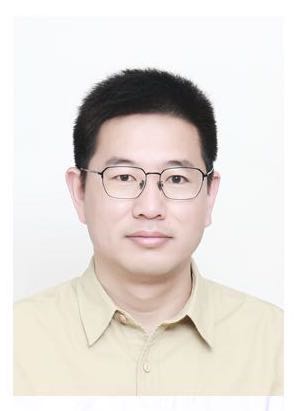}}]
	{Heng Li}{\space}
	received the B.S. degree from Mechanical Engineering, Hebei University of Technology, Tianjin, China, in 2010, and the M.S. degree from agricultural robotics technology, China Agricultural University, Beijing, China, in 2013. He is currently an engineer at Micro-Satellite Research Center of Zhejiang University, Hangzhou, China. His main research interest are software performance optimization of attitude and orbit control system.
\end{IEEEbiography}

\begin{IEEEbiography}
	[{\includegraphics[width=1in,height=1.25in,clip,keepaspectratio]{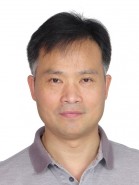}}]
	{Zhonghe Jin}{\space}
	received the Ph.D. degree from major of microelectronics and solid electronics, Zhejiang University, Hangzhou, China, in 1998. He is now the deputy director of Zhejiang University and the director of the center for micro satellite research. His main research areas include micro-satellites, MEMS etc.
	
\end{IEEEbiography}

\end{document}